\documentclass[12pt]{iopart}

\usepackage[squaren]{SIunits}
\usepackage{stmaryrd}
\usepackage{amsthm}
\usepackage{amssymb} % provides \Box
\usepackage{mathrsfs}
\usepackage{multirow}
\usepackage{color}
\usepackage{setspace} % provides spacing

\usepackage{graphicx} % provides includegraphics
\usepackage{subfig} % provides subfloat

\usepackage{url} % provides @online ou \url for bibtex

\usepackage{hyperref} % Hyperrefs and pdf settings

\begin{document}

\title[Hairy rotating black holes in cubic Galileon theory]{Hairy rotating black holes in cubic Galileon theory}
\author{K.~Van Aelst$^{1}$, E.~Gourgoulhon$^{1}$, P.~Grandcl\'ement$^{1}$, C.~Charmousis$^{2}$}
\vspace{10pt}
\begin{indented}
\item[$^{1}$] Laboratoire Univers et Th\'eories, Observatoire de Paris, Universit\'e PSL, CNRS, Universit\'e de Paris, 92190 Meudon, France
\item[$^{2}$] Laboratoire de Physique Th\'eorique, CNRS, Universit\'e Paris-Sud, Universit\'e Paris-Saclay, 91405 Orsay, France
\end{indented}
\ead{karim.van-aelst@obspm.fr, eric.gourgoulhon@obspm.fr, philippe.grandclement@obspm.fr, christos.charmousis@th.u-psud.fr}

\begin{abstract}
Numerical solutions for asymptotically flat rotating black holes in the cubic Galileon theory are presented.
These black holes are endowed with a nontrivial scalar field and exhibit a non-Schwarzschild behaviour:
faster than~$1/r$ convergence to Minkowski spacetime at spatial infinity and hence vanishing of the
Komar mass.
The metrics are compared with the Kerr metric for various couplings and angular velocities.
Their physical properties are extracted and show significant deviations from the Kerr case.
\end{abstract}


\section{Introduction}

Increasingly strong regimes of gravity are tested by modern, highly accurate instruments.
Over the last five years, the detectors of the LIGO/Virgo collaboration \cite{LIGO_Virgo_catalog_O1_O2},\nocite{GW_first_detection, GW170817, GW170817_and_counterpart}
the instrument GRAVITY \cite{GRAVITY}\nocite{GRAVITY_redshift_S2, GRAVITY_motion_ISCO}
and the Event Horizon Telescope \cite{EHT}\nocite{EHT_Shape_SgrA, EHT_Shadow_M87}
have collected data from objects involved in high energy gravitational processes:
coalescing compact objects \cite{GW_first_detection, GW170817},
stars and flares orbiting \emph{Sgr A*} \cite{GRAVITY_redshift_S2, GRAVITY_motion_ISCO},
accretion disks and shadows of supermassive black holes \cite{EHT_Shape_SgrA, EHT_Shadow_M87}.
So far, all these observations are consistent with the black hole model of general relativity (GR), adding to the successes of the latter in weaker gravitational regimes \cite{pulsar_timing,Will_review_tests_GR}.

Yet, many alternative theories of gravitation are being investigated \cite{review_Clifton, review_Berti, review_Nojiri_cosmo, review_Koyama_cosmo_tests} and this is important for at least two reasons.
Firstly, identifying all the modifications that lead to theoretical pathologies or observational incompatibilies is a relevant approach to understand better why GR is successful.
Secondly, GR actually suffers from several shortcomings or unresolved questions: on galactic and cosmological scales, it does not provide satisfactory explanations to the issues of dark matter and dark energy \cite{report_Saltas_cosmo, review_Nojiri_cosmo}, while it is expected to break down in the high energy or strong curvature regimes in view of its inadequacy to unify with the other fundamental interactions \cite{Kiefer_quantum_gravity, gravity_unification_review, review_Berti}.

Modified gravity theories aim to provide answers or alternative solutions to such questions or shortcomings of GR in the ultra-violet and infra-red sectors of gravity.
But then, the actual applicability of any modified theory of gravity is assessed from its compatibility with existing observational constraints and theoretical viability, e.g. well-posedness and stability.
For example, regarding observational constraints in the dark energy sector, the gravitational wave detection GW170817 and its electromagnetic counterpart GRB170817A set uptight constraints on the speed of gravitational waves \cite{GW170817_and_counterpart} (see also \cite{deRham:2018red} for a critical approach on the interpretation of these constraints).
Consequently, only restricted families of many modified theories of gravity turned out to be explicitly compatible with these constraints \cite{DE_after_GW_ahead}.
Regarding the theoretical analysis, the so-called ``no-hair'' theorems \cite{scalar_hair_review, hair_review} provide another kind of argument to appraise the relevance of a given modified theory.
Such theorems state the equivalence between a modified theory and GR relative to black hole solutions.
In other terms, these theorems single out conditions under which the black holes of a modified theory are as ``hairless'' as those of GR, i.e. belong to the Kerr-Newman family.
A typical example is that of Brans-Dicke gravity that has identical black hole solutions to GR (see \cite{Hawking_no_hair} and more recently \cite{no_hair_Sotiriou_1} and references within).

Horndeski theories are the most general scalar-tensor theories leading to second-order field equations, which, as shown in \cite{Kobayashi:2011nu}, coincide with the generalized covariant Galileon in four dimensions \cite{Galileon_original, covariant_Galileon, generalized_Galileons, review_Galileon}).
Within this context, the cubic Galileon theory is of particular interest among Horndeski theories \cite{Horndeski_original}.  For a start, it is the simplest of Galileons with higher order derivatives.
The cubic Galileon is also well-known for being related to the Dvali-Gabadadze-Porrati (DGP) braneworld model \cite{DGP_original}, from which all (flat) Galileon theories originate \cite{Galileon_original,review_Galileon,review_Galileons_de_Rham}.
More precisely, the DGP model is a 5-dimensional theory of gravity such that all non-gravitational fields are restricted to a 4-dimensional subspace (the usual spacetime), on which gravity is induced by a continuum of massive gravitons \cite{review_massive_de_Rham}.
In this framework, an effective formulation of gravity on the 4-dimensional spacetime generates the scalar term corresponding to the cubic Galileon theory in the decoupling limit \cite{DGP_effective_action_2}.
The DGP term, along with other covariant Galileons, also arises from Kaluza-Klein compactification of higher dimensional metric theories of gravity (see for example \cite{VanAcoleyen:2011mj, Charmousis:2014mia}).

On the observational side, the cubic Galileon is compatible with the observed speed of gravitational waves \cite{GW_speed, DE_after_GW_Vernizzi, DE_after_GW_ahead, DE_after_GW_revisited}.
% Referee 1: remark 3 and 4
Regarding cosmology, the cubic Galileon enters the family of theories featuring ``kinetic gravity braiding'' \cite{KGB}, which inherit infrared modifications of gravity from the DGP model.
These provide self-accelerating scenarios whose cosmological viability has been investigated in several studies, either assuming convergence of the Galileon to a common ``tracker'' solution \cite{Galileon_obs_status_Planck, Galileon_ISW_CMB} or more agnostic scenarios \cite{Galileon_recent_cosmo_data, Galileon_obs_status_cosmo_GW, Galileon_lensing_voids}.
These analyses highlighted strong tensions between the dark energy models of the cubic Galileon and observational data including e.g. the ISW effect.
Note though that the standard $\Lambda$CDM model may be recovered in the cubic Galileon, in which case these conclusions do not apply, in particular when the canonical kinetic term is not included.
% End of "Referee 1: remark 3 and 4"

On the theoretical side, various issues have been tackled within the framework of the cubic Galileon theory or larger theories including it:
accretion onto a black hole \cite{Galileon_accretion, stability_Galileon_accretion},
types of coupling to matter \cite{Galileon_cosmo_paths},
laboratory tests \cite{Galileon_labo},
cosmological dynamics \cite{cosmo_self_tun,cubic_cosmo}, structure formation \cite{cubic_Galileon_structure_formation},
stability of cosmological perturbations \cite{Galileon_cosmo_viability, Galileon_linear_cosmo_perturb},
well-posedness \cite{well_posedness_Love_Horn, stability_cosmo, well_posed_cubic_Horn, global_solutions_Horn}.
Finally, it was found in \cite{no_hair_Galileon} (with important  precisions given in \cite{hairy_BH_GB_Sotiriou_1, hairy_BH_GB_Sotiriou_2, Babichev:2016rlq}) that shift-symmetric Horndeski theory along with the cubic Galileon is subject to a no-hair theorem in the static and spherically symmetric case (see also \cite{slowly_rotating_no_hair} for an extension to slow rotation and \cite{Lehebel:2017fag} for stars).
This could have removed the interest for black holes in this theory, but instead it was rapidly shown that slightly violating one of the hypotheses of the no-hair of \cite{no_hair_Galileon}, namely the stationarity of the
scalar field \cite{Babichev:2013cya}, allowed to obtain static and spherically symmetric black holes different from GR solutions \cite{cubic_BH}.
This indicated that rotating black holes in the cubic Galileon theory might significantly deviate from the Kerr solution, which motivated the work reported here.

In fact, several other cases, constructed by breaking one of the hypotheses of the no-hair theorem, were found for different sectors of Horndeski theory and beyond (see for example \cite{hairy_BH_GB_Sotiriou_1, hairy_BH_GB_Sotiriou_2, Babichev:2017guv}).
Although these hairy solutions, with non trivial scalar field, are obtained for different higher order Horndeski terms, they can be separated in two generic classes: those in which spacetime is very close to that of GR, characterized by an additional parity symmetry of the action for the scalar ($\phi \leftrightarrow -\phi$) and often dubbed as stealth solutions; and those with no parity symmetry and significant departures from GR metrics.
For the former case a rotating stealth black hole was recently analytically constructed \cite{Charmousis:2019vnf}  making use of an analogy with geodesic congruences of Kerr spacetime \cite{Carter:1968rr}.
In the latter class, on the other hand, belong the DGP and the Gauss-Bonnet black holes (see for example the recent works \cite{Antoniou:2017hxj}, \cite{rotating_EdGB_1} and references within).
In this paper we will concentrate on the case of rotating black holes for the DGP Galileon finding significant deviations from the GR Kerr spacetime.

The structure of the paper is as follows.
Firstly, the field equations of the cubic Galileon theory are introduced in section \ref{section_dynamics} below.
Based on these equations, the no-scalar-hair theorem which the cubic Galileon is subject to, and the minimal way to circumvent it, are reviewed in section \ref{section_no_hair}.
This method provided the ansatz used for the scalar field in the rotating case.
It is described in section \ref{section_ansatze} along with the circular ansatz used for the metric.
The rest of the numerical setup is presented in sections \ref{section_equations} to \ref{section_accuracy}.
The numerical solutions are presented and analysed in sections \ref{section_num_sols} and \ref{section_phys}.

\section{The cubic Galileon model}
\label{section_Galileon}

\subsection{Dynamics}
\label{section_dynamics}

The vacuum action of the cubic Galileon involves the Einstein-Hilbert term (with a cosmological constant~$\Lambda$) and the usual scalar kinetic term for the scalar field~$\phi$ along with an additional nonstandard term:
\begin{eqnarray}
\label{eq_action}
S\left[ g,\phi \right]
&= \int \left[
            % \zeta (R^{(g)} - 2 \Lambda)
            \zeta (R - 2 \Lambda)
            - \eta (\partial \phi)^{2}
            + \gamma (\partial \phi)^{2} \Box\phi
        \right]
    \sqrt{\vert \det g \vert} d^{4}x,
\end{eqnarray}
where~$(\partial \phi)^{2} \equiv \nabla_{\mu} \phi \nabla^{\mu} \phi$ and~$\zeta$, $\eta$ and~$\gamma$ are coupling constants.

The scalar part of~(\ref{eq_action}) is known to emerge from an effective formulation of the DGP model \cite{DGP_original}.
In the DGP model, the usual spacetime is a timelike hypersurface of a 5-dimensional spacetime on which a metric alone is defined.
But the effective dynamics of gravity on the 4-dimensional spacetime involves scalar terms including those appearing in~(\ref{eq_action}) \cite{DGP_effective_action}, which actually become the only relevant contributions in some physically consistent decoupling limit \cite{DGP_effective_action_2}.

% Referee 2: sole remark (part 1/3); Referee 1: remarks 3 and 4
As an additional legacy from the DGP model, the cubic Galileon is subject to the Vainshtein mechanism \cite{original_Vainshtein,review_Vainshtein}, like all Galileon models which were originally designed to possess this property \cite{Galileon_original}.
This mechanism is based on nonlinear terms of the scalar Lagrangian that screen the non GR degrees of freedom on scales smaller than a certain ``Vainshtein'' radius around a spherical matter source.
It has been studied in different contexts such as massive gravity \cite{Babichev_Vainshtein_massive_gravity,Babichev_Vainshtein_bigravity} and Galileons \cite{Vainshtein_Galileons,Vainshtein_cubic_Galileon}.
For instance in the cubic Galileon theory, the dimension of the Solar System is smaller than the Vainshtein radius of the Sun, below which GR is recovered.
Hence in generic situations, local Solar System experiments and PPN methods cannot set constraints on the parameters of the theory \cite{PPNV_cubic}.

For the case of the cosmological galileon however, note that there are some subtleties due to kinetic gravity braiding, which is the fact that both scalar and metric equations involve second-derivatives of both~$g$ and~$\phi$ in any conformal frame (see for instance cubic Galileon equations~(\ref{eq_metric}) and~(\ref{eq_scalar}) below).
More precisely, the higher order nature of the Galileon operators, and in particular the presence of curvature in the scalar field equation, can invoke local constraints as explained in the careful analysis of \cite{cubic_cosmo}.
Yet these are evaded in the framework in which the work exposed below is set, notably due to asymptotic flatness (see sections \ref{section_BC} and \ref{section_stat_sphe}).
% End of "Referee 2: sole remark (part 1/3); Referee 1: remarks 3 and 4"

Explicitly, the metric equations in the cubic Galileon theory take the form
\begin{eqnarray}
\label{eq_metric}
G_{\mu\nu} + \Lambda g_{\mu\nu} = 8\pi T^{(\phi)}_{\mu\nu}
\end{eqnarray}
where
\begin{eqnarray}
\label{eq_T_phi_munu}
8\pi T^{(\phi)}_{\mu\nu} =
& \frac{\eta}{\zeta} \left( \partial_{\mu}\phi\partial_{\nu}\phi - \frac{1}{2}g_{\mu\nu}(\partial \phi)^{2} \right)
     \nonumber \\
& \phantom{(}
    + \frac{\gamma}{\zeta}
    \left(
            \partial_{(\mu}\phi\partial_{\nu)}(\partial \phi)^{2} - \Box\phi \partial_{\mu}\phi\partial_{\nu}\phi
            - \frac{1}{2}g_{\mu\nu}\partial^{\rho}\phi\partial_{\rho}[(\partial \phi)^{2}]
    \right)
\end{eqnarray}
does contain second derivatives of~$\phi$.

The scalar field equation actually coincides with the current conservation associated with the shift-symmetry~$\phi \rightarrow \phi + constant$
\footnote{Such symmetry is a remnant of the more general ``Galilean'' symmetry enjoyed by the action~(\ref{eq_action}) on Minkowski space \cite{Galileon_original}: $\phi \rightarrow \phi + constant,\ \nabla\phi \rightarrow \nabla\phi + constant\ vector$.}
of action~(\ref{eq_action}):
\begin{eqnarray}
\label{eq_scalar}
\nabla_{\mu} J^{\mu} = 0,
\end{eqnarray}
where
\begin{eqnarray}
\label{eq_current}
J_{\mu} &= \partial_{\mu}\phi \left( \gamma \Box\phi - \eta \right) - \frac{\gamma}{2} \partial_{\mu}\left(\partial\phi\right)^2,
\end{eqnarray}
which does generate second derivatives of the metric in~(\ref{eq_scalar}).

\subsection{No-scalar-hair theorem and hairy solutions}
\label{section_no_hair}

One of the first no-scalar-hair theorems was proven by Hawking in 1972 for stationary black holes in Brans-Dicke theory \cite{Hawking_no_hair}.
It was extended to a larger family of scalar-tensor theories by V. Faraoni and T. P. Sotiriou in 2012 using asymptotic flatness \cite{no_hair_Sotiriou_1, no_hair_Sotiriou_2}.
The result by L. Hui and A. Nicolis established in 2013 applies to another large family of scalar-tensor theories (shift-symmetric covariant Galileons) intersecting the former and including the cubic Galileon \cite{no_hair_Galileon}.
But this theorem applies to a case more restricted than stationarity and asymptotic flatness, as is reviewed below.

Prior to this, one can see from the field equations~(\ref{eq_metric}) and~(\ref{eq_scalar}) that any solution of vacuum GR along with a constant scalar field\footnote{This is equivalent to cancel everywhere due to the shift-symmetry of the theory.} is a solution to the cubic Galileon theory (see \cite{classification_Sotiriou} for general results on the theories featuring this property and their relations with other shift-symmetric theories).

The no-scalar-hair theorem stated in \cite{no_hair_Galileon} establishes the converse result in the case of an asymptotically flat, static, spherically symmetric black hole metric and a scalar field featuring the same symmetries and a standard kinetic term (i.e.~$\eta \neq 0$ in~(\ref{eq_action})): under such hypotheses, the solutions to the cubic Galileon theory can only be those of GR with a constant scalar field.
The proof, and an extension to the case~$\eta = 0$ (relevant for the work presented here, as detailed in section \ref{section_BC}), are given in \ref{appdx_no_hair} in the restricted case of the cubic Galileon.

Yet the attractiveness of a given modified theory is to feature deviations away from GR at least in some circumstances, otherwise there would be no interest in studying its black holes.
The solutions exhibited in \cite{cubic_BH} showed that such deviations exist in the cubic Galileon theory whenever the staticity of the scalar field is replaced by a linear time-dependence:
\begin{eqnarray}
\label{eq_scalar_ansatz_sphe}
\phi = qt + \Psi(R),
\end{eqnarray}
where $q$ is a constant, $t$ a time coordinate and $R$ a radial coordinate.
The structure~(\ref{eq_scalar_ansatz_sphe}) actually arises in a cosmological context from the assumption of a slow cosmological dynamics \cite{time_variation_qt}, and it has been considered in several contexts \cite{Galileon_accretion, cubic_cosmo, cosmo_self_tun} due to the following interesting properties.

% Referee 3: sole remark
Recall that the scalar field only contributes to the cubic Galileon action through its derivatives (hence the shift-symmetry).
As a result, the linear time-dependence of~(\ref{eq_scalar_ansatz_sphe}) does not bring any actual time-dependence into action~(\ref{eq_action}) and the field equations~(\ref{eq_metric}) and~(\ref{eq_scalar}), in which only the constant~$q$ appears.
This explains why Ansatz~(\ref{eq_scalar_ansatz_sphe}) is harmless in regard of instabilities that generically come with ever-growing fields: the perpetual increase with time cannot appear in any physical quantity.
Moreover, it is thus rigorously possible for the metric to be static and spherically symmetric and yet dressed with a scalar field not sharing all these symmetries.
% End of "Referee 3: sole remark"

Furthermore, the ansatz~(\ref{eq_scalar_ansatz_sphe}) does not spoil the self-consistency of the field equations in the static and spherically symmetric case; this means that one is left with as many unknown functions as independent ordinary differential equations \cite{cubic_BH}, suited for numerical integration by a shooting method.
Last but not least, it has been shown for cases where analytical expressions are known, \cite{Babichev:2013cya,Babichev:2016rlq}, that the linear time dependence (\ref{eq_scalar_ansatz_sphe}) renders the scalar field regular at the event horizon by precisely cancelling out the radial divergence in $\Psi(R)$.
The existence of black hole solutions different from GR provided a path to hairy rotating solutions, whose numerical construction is now presented.

\section{Numerical setup}
\label{section_num_setup}

\subsection{Ans\"{a}tze and assumptions}
\label{section_ansatze}

The goal is to construct stationary, rotating (i.e. axisymmetric with a nonzero angular velocity), asymptotically flat black hole spacetimes.
In addition, a simplifying assumption is made: the spacetime geometric structure is assumed to be circular, or ``$t,\varphi$-orthogonal'' (see \cite{Carter_killing_ortho_transitive,Carter_Houches,heusler_uniqueness_book,intro_relat_stars,Chandra_book,Eric_Silvano_noncircular} for further details on the statements reported in this section).
The accuracy of this hypothesis will be evaluated in section \ref{section_accuracy}.

Denoting~$\xi$ and~$\chi$ the Killing vectors associated with stationarity and axisymmetry respectively, circularity amounts to requiring that there exists a coordinate system~$(t, x^{1}, x^{2}, \varphi)$ such that~$\xi = \partial_{t}$,~$\chi = \partial_{\varphi}$ and the transformation~$(t, \varphi) \mapsto (-t, -\varphi)$ leaves the metric unchanged.
This is equivalent to complete integrability of the codistribution~$(dt, d\varphi)$, i.e. the existence of a foliation of spacetime by 2-surfaces (called meridional surfaces) everywhere orthogonal to~$\xi$ and~$\chi$.
Using Frobenius theorem, this property takes the form
\begin{eqnarray}
\label{eq_Frob_circu}
d\xi \wedge \xi \wedge \chi = d\chi \wedge \xi \wedge \chi = 0,
\end{eqnarray}
where the vectors are identified with their corresponding 1-form by metric duality.

Since the surfaces of transitivity (i.e. the orbits of the combined actions of~$\xi$ and~$\chi$) are orthogonal to the meridional surfaces, the metric components~$(t x^{1})$,~$(t x^{2})$,~$(\varphi x^{1})$ and~$(\varphi x^{2})$ vanish in coordinate systems having the aforementioned properties.
A judicious choice of coordinates~$(x^{1}, x^{2})$ within the meridional surfaces allows to cancel~$g_{x^{1}x^{2}}$ as well so that the metric reads\footnote{When such a choice is made, $x^{1}$ and $x^{2}$ are rather denoted~$r$ and~$\theta$ respectively.}
\begin{eqnarray}
\label{eq_circular}
ds^{2} = - N^2 dt^2 + A^2 \left( dr^2 + r^2 d\theta^2 \right) + B^2 r^2 \sin^2 \theta \left( d\varphi - \omega dt \right)^2,
\end{eqnarray}
where~$N$,~$A$,~$B$ and~$\omega$ are only functions of the coordinates~$r$ and~$\theta$.
Such a coordinate system is naturally called quasi-isotropic.
In the case of spherical symmetry, the four functions only depend on~$r$, while~$\omega = 0$ and~$A = B$ (so that the coordinates are merely called isotropic).

In a circular spacetime, Ricci-circularity holds, i.e.
\begin{eqnarray}
\label{eq_Ric_circu}
Ric(\xi) \wedge \xi \wedge \chi = Ric(\chi) \wedge \xi \wedge \chi = 0,
\end{eqnarray}
where~$Ric$ is the Ricci tensor.
In stationary, axisymmetric, asymptotically flat spacetimes, the converse result is true, i.e.~(\ref{eq_Ric_circu})~$\Rightarrow$~(\ref{eq_Frob_circu}).
Then, within GR, the Einstein equations allow to substitute the Ricci tensor with the energy-momentum tensor~$T$\footnote{Because one always has $g(\xi) \wedge \xi \wedge \chi = g(\chi) \wedge \xi \wedge \chi = 0$.}, so that an asymptotically flat black hole is circular if and only if the following holds (generalized Papapetrou theorem):
\begin{eqnarray}
\label{eq_T_circu}
T(\xi) \wedge \xi \wedge \chi = T(\chi) \wedge \xi \wedge \chi = 0.
\end{eqnarray}

This indicates that circularity may be interpreted in terms of the physical dynamics of matter rather than purely geometric statements.
More precisely, the relations~(\ref{eq_T_circu}) indicate that the source of the gravitational field has purely rotational motion about the symmetry axis and no momentum currents in the meridional planes.
Hence assuming circularity is very standard in numerical relativity to handle rapidly rotating stars since such objects have negligible convective meridional flows compared to rotation-induced circulation \cite{circular_rotating_relativistic_bodies}.
For instance, circularity allowed to model rotating proto-neutron stars in general relativity \cite{circular_rotating_proto_neutron_stars}.
In the case of a scalar field, circular rotating boson stars were also constructed numerically \cite{circular_rotating_boson_star}.
Finally, circularity is very relevant to describe rotating black holes: the Kerr solution is circular\footnote{The transformation from the usual Boyer-Lindquist coordinates to quasi-isotropic coordinates can be established explicitly \cite{quasi_isotropic_Kerr}.} and numerical metrics of rotating black holes were successfully computed in Einstein-Yang-Mills theory \cite{Einstein_SU2YM_1,Einstein_SU2YM_2} and in the dilatonic Einstein-Gauss-Bonnet theory \cite{rotating_EdGB_1,rotating_EdGB_2} based on circularity.

Regarding the scalar field, the successful ansatz~(\ref{eq_scalar_ansatz_sphe}) is rehashed, with a mere additional angular dependence, in order to connect with the solutions of \cite{cubic_BH} in the non-rotating limit:
\begin{eqnarray}
\label{eq_scalar_ansatz}
\phi = qt + \Psi(r,\theta).
\end{eqnarray}

\subsection{Equations in quasi-isotropic gauge}
\label{section_equations}

Injecting the ans\"{a}tze~(\ref{eq_circular}) and~(\ref{eq_scalar_ansatz}) into the metric equations~(\ref{eq_metric}) yields eight nontrivial equations rather than ten since the components~$(r,\varphi)$ and~$(\theta,\varphi)$ of the three tensors appearing in~(\ref{eq_metric}) all separately vanish.
These eight metric equations are combined to form four coupled, independent equations adding to the scalar equation (\ref{eq_scalar}) to solve for the four metric functions~$N$, $A$, $B$, $\omega$ and the scalar function~$\Psi$.

Every quantity is then made dimensionless using the free parameters of the theory, which are the scalar velocity~$q$, the cosmological constant~$\Lambda$, the coupling constants~$\zeta$, $\eta$ and $\gamma$, and the event horizon radial coordinate~$r_{\mathcal{H}}$ (in quasi-isotropic coordinates, the event horizon is always located at a constant radial coordinate):
\begin{eqnarray}
\bar{\Lambda} \equiv \Lambda r_{\mathcal{H}}^{2}, \hspace{0.1\textwidth}
&\bar{\eta} \equiv -q^{2} r_{\mathcal{H}}^{2} \frac{\eta}{\zeta}, \hspace{0.1\textwidth}
&\bar{\gamma} \equiv q^{3} r_{\mathcal{H}} \frac{\gamma}{\zeta}, \\
\bar{r} \equiv \frac{r}{r_{\mathcal{H}}},
&\bar{\omega} \equiv r_{\mathcal{H}} \omega,
&\bar{\Psi} \equiv \frac{\Psi}{q r_{\mathcal{H}}},
\end{eqnarray}
and all the functions are manipulated as functions of~$\bar{r}$.

Eventually, the four metric equations schematically take the form
\begin{eqnarray}
N^{2} \Delta_{3} N = \mathcal{S}_{N}, \label{eq_metric_QI_N} \\
N^{3} \Delta_{2} [NA] = \mathcal{S}_{A}, \label{eq_metric_QI_NA} \\
N^{2} \Delta_{2} [NB\bar{r}\sin\theta] = \mathcal{S}_{B}, \label{eq_metric_QI_NB} \\
N \Delta_{3} [\bar{\omega} \bar{r} \sin \theta] = \mathcal{S}_{\bar{\omega}}, \label{eq_metric_QI_adom}
\end{eqnarray}
where the right-hand side terms are explicitly given in \ref{appdx_RHS} and the following notations are used:
\begin{eqnarray}
\Delta_{2} = \partial^{2}_{\bar{r}\bar{r}} + \frac{1}{\bar{r}} \partial_{\bar{r}} + \frac{1}{\bar{r}^{2}} \partial^{2}_{\theta\theta}, \\
\Delta_{3} = \partial^{2}_{\bar{r}\bar{r}} + \frac{2}{\bar{r}} \partial_{\bar{r}} + \frac{1}{\bar{r}^{2}} \partial^{2}_{\theta\theta} + \frac{1}{\bar{r}^{2} \tan\theta} \partial_{\theta}, \\
\tilde{\Delta}_{3} = \Delta_{3} - \frac{1}{\bar{r}^{2} \sin^{2}\theta}.
\end{eqnarray}

Once again, recall that the cubic Galileon theory features the shift-symmetry $\phi \rightarrow \phi + constant$, meaning that only the first derivatives of~$\phi$ are physically meaningful.
The numerical approach presented in section \ref{section_numeric} below concretely makes use of this fact: within the numerical code, the scalar field is only manipulated through its first derivatives~$\bar{\Psi}' \equiv \partial_{\bar{r}} \bar{\Psi}$ and~$\bar{\Psi}_{\theta} \equiv \partial_{\theta} \bar{\Psi}$.
More precisely, $\bar{\Psi}'$ and $\bar{\Psi}_{\theta}$ are first introduced as independent functions, just like~$N$, $A$, $B$ and $\bar{\omega}$.
The fact that these functions actually arise from a common scalar field is then implemented through imposing~$\partial_{\theta} \bar{\Psi}' = \partial_{\bar{r}} \bar{\Psi}_{\theta}$ (symmetry of second-derivatives) in addition to the equations~$\{(\ref{eq_metric_QI_N})-(\ref{eq_metric_QI_adom}), (\ref{eq_scalar})\}$.

The complete set of equations to solve then is
\begin{eqnarray}
N^{2} \Delta_{3} N = \mathcal{S}_{N}, \label{eq_metric_QI_N_bis} \\
N^{3} \Delta_{2} [NA] = \mathcal{S}_{A}, \label{eq_metric_QI_NA_bis} \\
N^{2} \Delta_{2} [NB\bar{r}\sin\theta] = \mathcal{S}_{B}, \label{eq_metric_QI_NB_bis} \\
N \Delta_{3} [\bar{\omega} \bar{r} \sin \theta] = \mathcal{S}_{\bar{\omega}}, \label{eq_metric_QI_adom_bis} \\
\partial_{\theta} \bar{\Psi}' = \partial_{\bar{r}} \bar{\Psi}_{\theta} \label{eq_Schwarz}, \\
\nabla_{\mu} \bar{J}^{\mu} = 0, \label{eq_scalar_bis}
\end{eqnarray}
where the explicit expression of the scalar equation~(\ref{eq_scalar_bis}) is also given in \ref{appdx_RHS}.

Of course, if a circular black hole exists in the cubic Galileon theory, then it satisfies the system~(\ref{eq_metric_QI_N_bis})-(\ref{eq_scalar_bis}).
But any solution to this system does not necessarily satisfy all the metric equations of motion~(\ref{eq_metric}) since only four independent combinations of the latter are solved instead of eight.
Hence each numerical solution to~(\ref{eq_metric_QI_N_bis})-(\ref{eq_scalar_bis}) was reinjected into the whole set of metric equations~(\ref{eq_metric}) to assess the relevance of the circularity hypothesis a posteriori (see section \ref{section_accuracy}).

\subsection{Boundary conditions}
\label{section_BC}

Equations~(\ref{eq_metric_QI_N_bis})-(\ref{eq_scalar_bis}) form a system of first (equation~(\ref{eq_Schwarz})) and second order coupled partial differential equations (PDE) involving the six functions~$N$, $A$, $B$, $\bar{\omega}$, $\bar{\Psi}'$ and~$\bar{\Psi}_{\theta}$.
It must then be provided with boundary conditions suitable for the search for black hole solutions with nontrivial scalar hair.
More precisely, the system is defined on a meridional surface (all of them are equivalent due to circularity) between the intersections of the latter with the black hole event horizon and spacetime infinity.
As mentioned in section \ref{section_equations}, the event horizon is located at~$\bar{r} = 1$, while spacetime infinity corresponds with the limit~$\bar{r} \rightarrow \infty$.
Boundary conditions must then be prescribed for both limits.

First, in quasi-isotropic coordinates, the function~$N$ must vanish on the event horizon (see for instance \cite{quasi_isotropic_Kerr} for the case of Kerr).
This induces an important alteration of the nature of the equations~(\ref{eq_metric_QI_N_bis})-(\ref{eq_metric_QI_adom_bis}) since all the second-order operators acting on the metric functions thus cancel at~$\bar{r} = 1$.
This kind of degeneracy actually reduces the required number of boundary conditions.

The other crucial condition at the horizon is the value of the function~$\bar{\omega}$.
The weak rigidity theorem states the existence of a constant~$\Omega_{\mathcal{H}}$ such that~$\xi + \Omega_{\mathcal{H}} \chi$ is (a Killing vector field) normal to the horizon \cite{Carter_Houches, heusler_uniqueness_book}.
On the horizon, the function~$\bar{\omega}$ necessarily equals the constant~$\bar{\Omega}_{\mathcal{H}} \equiv r_{\mathcal{H}} \Omega_{\mathcal{H}}$, called the dimensionless angular velocity of the horizon.

Regarding conditions at infinity, the only case considered in this paper is asymptotic flatness.
% Referee 1: remarks 1 and 3 (and 4); Referee 2: sole remark (part2/3)
This is not meant to fit with observations of the Universe based on which a small positive value is credited to an effective cosmological constant, which is usually modeled by asymptotically de Sitter models such as the dark energy scenarios of the Galileons mentioned in the introduction.
Rather, the prime objective of the present work is to construct the strong-field region of rotating black holes in the simplest Galileon with higher-order derivatives.
This actually prepares for later investigation of their geodesics and the imaging of an emitting accretion torus surrounding them, which concerns scales much smaller than a potential cosmological horizon.
Furthermore, asymptotic flatness is a standard hypothesis made to study isolated black holes and establish no-hair theorems \cite{no_hair_Sotiriou_1}.
In particular, it leads to construct hairy black holes that escape the no-hair theorem of \cite{no_hair_Galileon} in a minimal way.

Yet, in the cubic Galileon theory, imposing asymptotic flatness in static and spherical symmetry is incompatible with Ansatz~(\ref{eq_scalar_ansatz_sphe}) unless~$\eta = \Lambda = 0$.
% End of "Referee 1: remarks 1 and 3 (and 4); Referee 2: sole remark (part2/3)"
To picture this, it is easier to consider the Schwarzschild-like coordinates $(t,R,\theta,\varphi)$ used in \cite{cubic_BH}, with respect to which the static and spherically symmetric line element takes the form
\begin{eqnarray}
\label{eq_metric_Schw_like}
ds^{2} = - h(R) dt^{2} + \frac{1}{f(R)} dR^{2} + R^{2} \left( d\theta^{2} + \sin^{2}\theta d\varphi^{2} \right).
\end{eqnarray}

Using the scalar ansatz~(\ref{eq_scalar_ansatz_sphe}), all the relevant equations are the following (the~$(tR)$ equation\footnote{In this context, the~$(tR)$ equation implies the scalar equation \cite{cubic_BH}.}, a combination of the~$(tR)$ and~$(RR)$ equations, and a combination of the~$(tR)$, $(RR)$ and~$(tt)$ equations respectively):
\begin{eqnarray}
\label{eq_4_1}
\gamma (R^{4}h)' fh \Psi'^{2} - \gamma q^{2} R^{4} h' - 2 \eta R^{4} h^{2} \Psi' = 0, \\
\label{eq_4_2}
\frac{\eta}{2\zeta} (fh\Psi'^{2} - q^{2}) + \frac{fh'}{R} + h \left( \frac{f-1}{R^{2}} + \Lambda \right)= 0, \\
\label{eq_4_3}
f\Psi'^{2} \left[ \eta R^{2} \sqrt{\frac{h}{f}} -\gamma \left( R^{2}\sqrt{fh} \Psi' \right)' \right] = 2\zeta Rh \left( \sqrt{\frac{f}{h}} \right)',
\end{eqnarray}
where a prime denotes differentiation with respect to the unique variable~$R$.

As mentioned in section \ref{section_no_hair}, one can note that the Schwarzschild-(Anti-)de Sitter metric along with~$\Psi' = 0$ and~$q = 0$ (i.e.~$\phi = constant$) must be a solution to the system~(\ref{eq_4_1})-(\ref{eq_4_3}) since it is a static and spherically symmetric vacuum solution of general relativity:
\begin{eqnarray}
\label{eq_Schw_AdS}
h(R) = f(R) = 1 - \frac{\mu}{R} - \frac{\Lambda}{3} R^2,
\end{eqnarray}
where~$\mu$ appears as an integration constant

According to \cite{cubic_BH}, injecting asymptotic expansions in powers of~$1/R$ for~$h$, $f$ and~$\Psi$ into~(\ref{eq_4_1})-(\ref{eq_4_3}) yields the following asymptotic behaviours if~$\eta \neq 0$:
\begin{eqnarray}
\label{eq_asympt_h}
h(R) = - \frac{\Lambda_{\rm eff}}{3} R^2 + 1 + O\left( \frac{1}{R} \right), \\
\label{eq_asympt_f}
f(R) = - \frac{\Lambda_{\rm eff}}{3} R^2 + c + O\left( \frac{1}{R} \right), \\
\label{eq_asympt_chi}
h(R) \Psi'(R) = \frac{\eta R}{3\gamma} + \frac{c'}{R} + O\left( \frac{1}{R^{2}} \right),
\end{eqnarray}
where~$c$ and~$c'$ are some fixed constants and~$\Lambda_{\rm eff}$ is an effective cosmological constant made from a combination of the bare cosmological constant~$\Lambda$ and the kinetic coupling~$\eta$.

Therefore, if~$\Lambda_{\rm eff} \neq 0$, spacetime is asymptotically (anti-)de Sitter.
Asymptotic flatness thus requires~$\Lambda_{\rm eff} = 0$, which is impossible whenever~$\eta \neq 0$ (see the relations (4.10) of \cite{cubic_BH}).

Then, setting~$\eta$ to~$0$ in~(\ref{eq_4_2}) yields
\begin{eqnarray}
\label{eq_4_2_eta0}
f \left( \frac{h'}{Rh} + \frac{1}{R^{2}} \right) = \frac{1}{R^{2}} - \Lambda,
\end{eqnarray}
while asymptotic flatness (i.e. vanishing Riemann tensor when~$R \rightarrow \infty$) requires the following asymptotic behaviours:
\begin{eqnarray}
\label{eq_asympt_flat_h}
\frac{h'}{h} = o\left( \frac{1}{R} \right),     \\
\label{eq_asympt_flat_f}
f \longrightarrow 1,
\end{eqnarray}
so that~$\Lambda$ must be~$0$ as well as~$\eta$.

As mentioned in section \ref{section_no_hair}, it is shown in \ref{appdx_no_hair} that, for the cubic Galileon, the no-hair theorem still holds if~$\eta = 0$.
Therefore, the asymptotically flat, static, spherically symmetric hairy solutions constructed in \cite{cubic_BH} with~$\eta = \Lambda = 0$ evade the no-hair theorem in a minimal fashion since only the staticity of the scalar field is abandoned.

It is reasonable to think that asymptotic flatness requires vanishing~$\eta$ and~$\Lambda$ even in the rotating case, although there is no proof of such a claim.
Regardless of the actual answer,~$\eta$ and~$\Lambda$ are set to zero in the numerical work exposed in this paper in order to connect with the solutions of \cite{cubic_BH} in the non-rotating limit.

\subsection{Numerical treatment}
\label{section_numeric}

The numerical approach to solve the above problem comprises two steps implemented within the library \emph{Kadath} \cite{Kadath}.
First, the system~(\ref{eq_metric_QI_N_bis})-(\ref{eq_scalar_bis}) is discretized within the framework of spectral methods.
This amounts to project each function~$N$, $A$, $B$, $\bar{\omega}$, $\bar{\Psi}'$ and $\bar{\Psi}_{\theta}$ onto a set of basis functions defined as the products of (Legendre or Chebyshev) polynomials~$T_{i}$ with trigonometric functions, e.g. for the function~$A$:
\begin{eqnarray}
A(r,\theta) = \sum_{i=0}^{m_{r}} \sum_{j=0}^{m_{\theta}} &\tilde{A}_{ij} T_{i}(r) \cos(2j\theta),
\end{eqnarray}
where~$m_{r}$ and~$m_{\theta}$ are integers defining the resolution of the discretization\footnote{For class~$C^{\infty}$ functions, the convergence of the spectral series towards the original function is exponential in the resolution.}.
All the information about the unknown function~$A$ is then encoded into the spectral coefficients~$\tilde{A}_{ij}$.
Moreover, the projection of any of its partial derivative is also given in terms of these coefficients.
Applying this procedure to each unknown function~$N$, $A$, $B$, $\bar{\omega}$, $\bar{\Psi}'$ and~$\bar{\Psi}_{\theta}$ in the system~(\ref{eq_metric_QI_N_bis})-(\ref{eq_scalar_bis}) transforms the latter into a nonlinear algebraic system~$S$, whose unknowns are suitable combinations of the spectral coefficients ensuring regularity conditions \cite{Kadath}.

Secondly, the discretized system~$S$ is solved with a Newton-Raphson algorithm.
The vector~$\tilde{X}$ gathering all the relevant combinations of the spectral coefficients should satisfy~$S(\tilde{X}) = 0$.
Starting with an initial guess~$\tilde{X}^{(0)}$ and denoting~$\tilde{X}^{(n)}$ the vector gathering the coefficients at step~$n$, $\tilde{X}^{(n+1)}$ is built as the solution to~$S(\tilde{X}^{(n)}) + dS_{\tilde{X}^{(n)}} ( \tilde{X}^{(n+1)} - \tilde{X}^{(n)} ) = 0$, which requires inverting the Jacobian matrix~$dS_{\tilde{X}^{(n)}}$.

Under appropriate conditions, such an iterative process converges towards the exact solution of~$S$.
In particular, a good initial guess is an important condition of success.
This merely means that the closest to the exact solution the process starts, the more chance it has to converge to the solution (while starting too far away from it induces risks to leave the neighbourhood of the solution after a few iterations and eventually diverge).
This is the reason why the existence of the static and spherically symmetric black hole solutions of \cite{cubic_BH} is particularly useful since reconstructing these solutions numerically (for a fixed choice of the coupling constants) provides ideal initial guesses to reach slowly rotating solutions, which in turn serve as initial guesses to reach slighlty more rapidly rotating solutions and so on.

\setcounter{footnote}{0} % solves the error "counter too large"
Finally, let us mention that, in the rotating cases, an additional condition was implemented in order to avoid a conical singularity \cite{intro_relat_stars}.
It consists in imposing~$A = B$ on the symmetry axis~$\theta = 0, \pi/2$\footnote{In the static and spherically symmetric cases,~$A$ spontaneously equals to~$B$ everywhere (as it should in spherical symmetry) through the numerical process without being imposed anywhere.}, which guarantees that the metric could be regularly well-defined on an open chart containing the axis.
For instance, this condition was also imposed in \cite{rotating_EdGB_1,rotating_EdGB_2} to construct rotating black holes in the dilatonic Einstein-Gauss-Bonnet theory, but for rotating bosons stars \cite{circular_rotating_boson_star}, the field equations alone imply~$A = B$ on the symmetry axis.

\subsection{Accuracy of the code}
\label{section_accuracy}

\begin{figure}
    \begin{center}
        \subfloat[Validation of equation~$E_{r\theta}$ (the error decreases with the resolution).]
        {
            \includegraphics[width=0.47\textwidth]{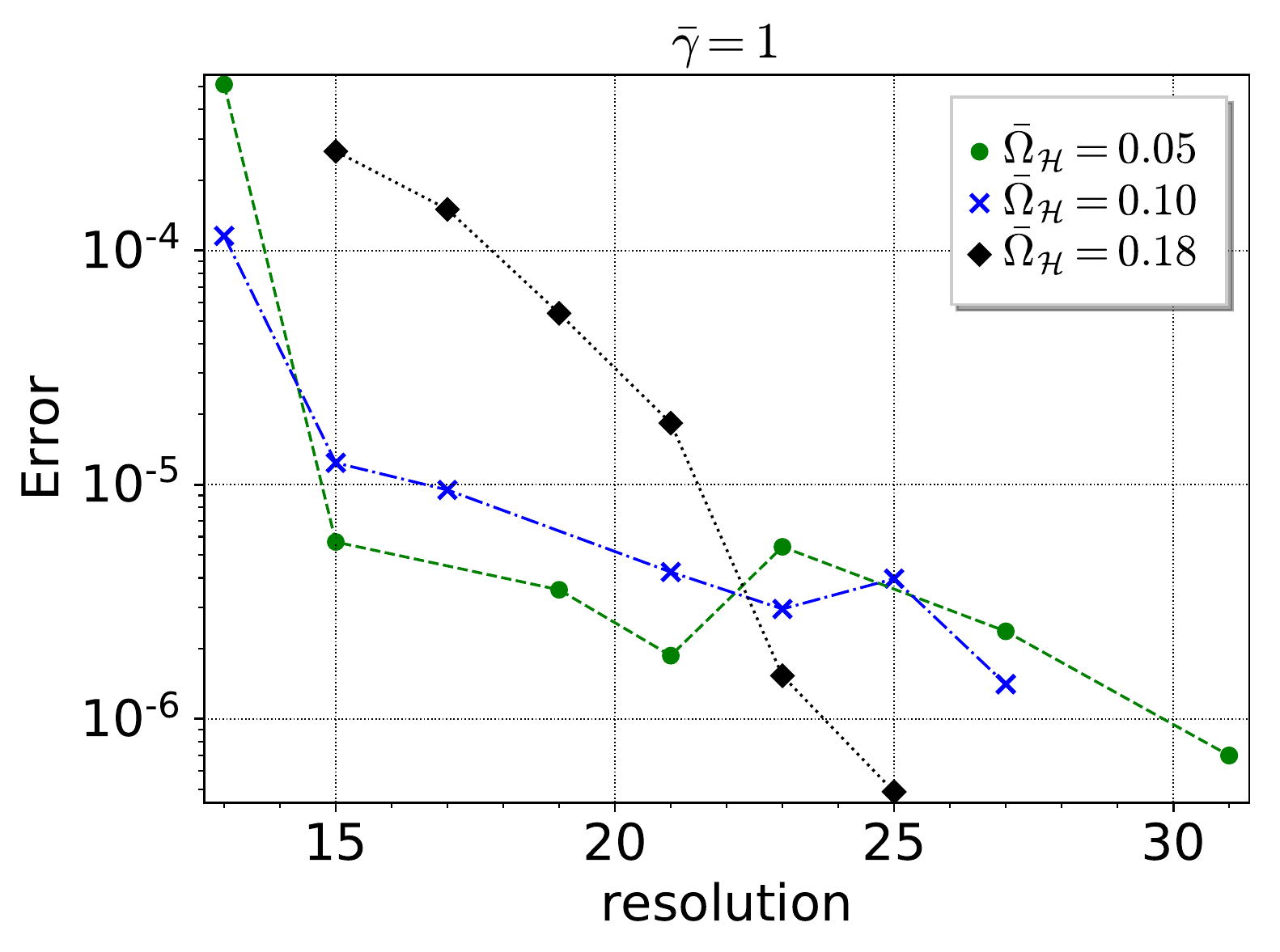}
            \label{fig_error_evol_21_fixed_adgamma}
        }
        \subfloat[Violation of equation~$E_{tr}$ (the error is independent of the resolution).]
        {
            \includegraphics[width=0.47\textwidth]{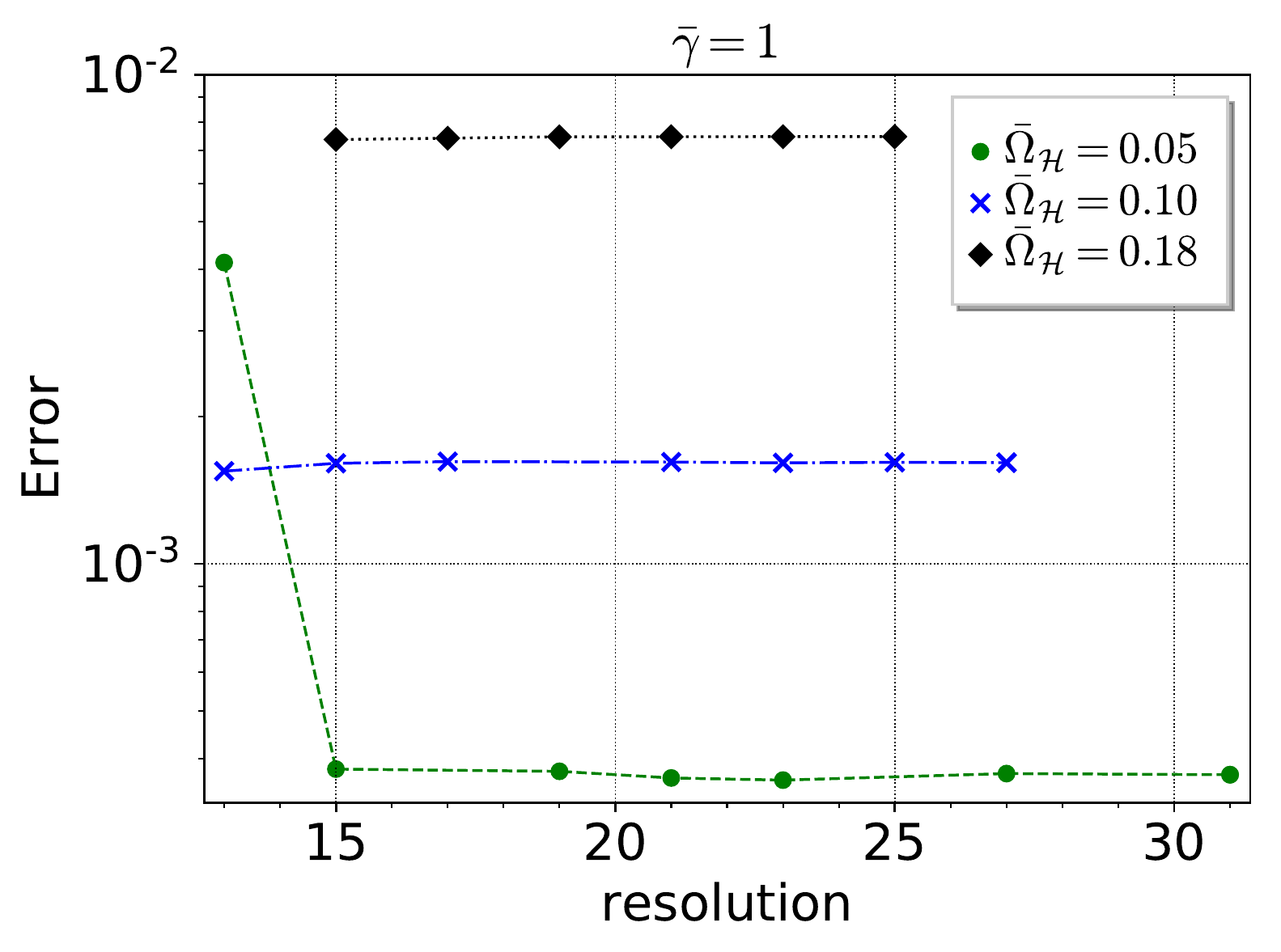}
            \label{fig_error_momentum_1_fixed_adgamma}
        }
    \end{center}
\caption{Errors on the metric equations with respect to the resolution $m_{r} = m_{\theta}$. The error is measured in horizon units~($\bar{r} = 1$).}
\label{fig_errors_metric_equations}
\end{figure}

As explained in section \ref{section_equations}, all the numerical solutions to the system~(\ref{eq_metric_QI_N_bis})-(\ref{eq_scalar_bis}) were reinjected into the whole set of metric equations~(\ref{eq_metric}) in order to assess the validity of the code.
Writing the metric equations~(\ref{eq_metric}) as~$E_{\mu\nu} = 0$, the error on each equation corresponds to its maximum spectral coefficent (in absolute value).
Six out of the eight nontrivial\footnote{See section \ref{section_equations}.} metric equations feature a fast decrease of the error as the resolution increases, which confirms that these equations are properly solved numerically.
Figure \ref{fig_error_evol_21_fixed_adgamma} illustrates this fact in the case of equation~$E_{r\theta}$ for various angular velocities at fixed coupling $\bar{\gamma} = 1$.

On the other hand, the error on the two metric equations~$E_{tr}$ and~$E_{t\theta}$ is independent of the resolution, as illustrated on Fig~\ref{fig_error_momentum_1_fixed_adgamma}, revealing that there exists an actual violation of non numerical origin.
The cause of this violation can be identified a bit more precisely.
In quasi-isotropic coordinates, the components~$(tr)$ and~$(t\theta)$ of both the metric and Ricci tensors are zero.
As a result, the metric equations~$E_{tr} = E_{t\theta} = 0$ reduce to~$T^{(\phi)}_{tr} = T^{(\phi)}_{t\theta} = 0$.
Actually, these last two equations coincide with the two nontrivial circularity conditions provided by the generalized Papapetrou theorem (\ref{eq_T_circu}).
The latter can be applied to~$T^{(\phi)}$ because the metric equations~$E_{\mu\nu} = 0$ have an Einstein-like structure.
This yields
\begin{eqnarray}
\label{eq_T_circu_radial}
\left( T^{(\phi)}(\partial_{t}) \wedge \partial_{t} \wedge \partial_{\varphi} \right)_{tr\varphi}
= T^{(\phi)}_{t[t} \left( \partial_{t} \right)_{r} \left( \partial_{\varphi} \right)_{\varphi]}
\propto T^{(\phi)}_{tr},    \\
\left( T^{(\phi)}(\partial_{t}) \wedge \partial_{t} \wedge \partial_{\varphi} \right)_{t\theta\varphi}
= T^{(\phi)}_{t[t} \left( \partial_{t} \right)_{\theta} \left( \partial_{\varphi} \right)_{\varphi]}
\propto T^{(\phi)}_{t\theta}.
\end{eqnarray}

Therefore the errors on~$E_{tr}$ and~$E_{t\theta}$ estimate the validity of the circularity hypothesis.
More precisely, in the expression (\ref{eq_T_phi_munu}) of~$T^{(\phi)}_{tr}$ (resp.~$T^{(\phi)}_{t\theta}$), the only nontrivial terms are those proportional to~$\partial_{t}\phi\partial_{r}\phi$ (resp.~$\partial_{t}\phi\partial_{\theta}\phi$) and~$\partial_{(t}\phi\partial_{r)}(\partial \phi)^{2}$ (resp.~$\partial_{(t}\phi\partial_{\theta)}(\partial \phi)^{2}$) which are nonzero only if~$\phi$ depends on both~$t$ and~$r$ (resp.~$t$ and~$\theta$).
This means that non circularity is caused by combined time and radial, or time and angular, dependences of the scalar field.
Yet the ansatz (\ref{eq_scalar_ansatz_sphe}) used in \cite{cubic_BH} to derive static and spherically symmetric solutions does feature both time and radial dependences.
But these solutions were obtained taking advantage of the fact that~$E_{tR}$ (in the Schwarzschild-like coordinates (\ref{eq_metric_Schw_like})) implies the scalar equation.
Thus, solving~$E_{tR} = 0$ instead of the scalar equation automatically fulfilled the circularity condition (\ref{eq_T_circu_radial}) since~$E_{tR} \propto T^{(\phi)}_{tR} \propto T^{(\phi)}_{tr}$ (where the last relation holds because the transformation (\ref{eq_change_Schw_QI}) from Schwarzschild-like coordinates to quasi-isotropic coordinates relates only the coordinates~$R$ and~$r$ in spherical symmetry).

But as soon as one looks for rotating solutions and thus adds an angular dependence to all functions, including the scalar field according to the ansatz (\ref{eq_scalar_ansatz}), the equations are too complex to benefit from a similar simplification.
Therefore the system~(\ref{eq_metric_QI_N_bis})-(\ref{eq_scalar_bis}) based on the circular metric (\ref{eq_circular}) and the ansatz (\ref{eq_scalar_ansatz}) is not exactly self-consistent.
Yet, the violation of circularity in the dimensionless setup is less than~$10^{-2}$, meaning that it is fairly small with respect to the scale given by the radial coordinate~$r_{\mathcal{H}}$ of the event horizon in a dimensional physical configuration.
In addition, Fig~\ref{fig_error_momentum_1_fixed_adgamma} expectedly confirms that the violation continuously goes to zero with the angular velocity (since in this limit the solutions are exactly circular), so that it seems reasonable to believe that the solutions presented in the next sections still provide precise approximations to rotating black hole solutions of the cubic Galileon theory.

\section{Black hole solutions}
\label{section_num_sols}

\subsection{Static and spherically symmetric black holes}
\label{section_stat_sphe}

First, the existing static, spherically symmetric black hole solutions reported in \cite{cubic_BH} have been reconstructed in the quasi-isotropic gauge (instead of the Schwarzschild-like coordinates used in \cite{cubic_BH}) in order to later serve as initial guesses to compute rotating solutions.
As mentioned in \ref{section_no_hair}, these solutions were obtained in \cite{cubic_BH} by numerical integration of the ordinary differential equations~$(\ref{eq_4_1})-(\ref{eq_4_3})$.
In addition, the value of~$h'$ was prescribed at the horizon in order to obtain the desired asymptotic behaviour (shooting method).
In this paper, these solutions are generated with the numerical treatment presented in section \ref{section_numeric}, i.e. as solutions to the PDE system~(\ref{eq_metric_QI_N_bis})-(\ref{eq_scalar_bis}).
In addition, boundary conditions are prescribed both at infinity and at the horizon; in particular, staticity is imposed by setting the dimensionless angular velocity~$\bar{\Omega}_{\mathcal{H}}$ to zero.
The resulting numerical solutions feature spherical symmetry ($A = B$,~$\bar{\omega} = 0$ everywhere, and no angular dependence) although such symmetry is not imposed anywhere in the numerical process.

As explained in section \ref{section_numeric}, the numerical process requires initial guesses.
Conveniently, the test-field solution given in \cite{cubic_BH} (relations (4.12)-(4.13)) provides the very first of them.
This configuration merely comes out from solving the scalar equation~(\ref{eq_scalar}) on a Schwarzschild background metric with the scalar ansatz~(\ref{eq_scalar_ansatz_sphe}), which physically amounts to neglecting the back-reaction of the scalar field onto the metric, i.e. taking the limit~$\gamma \rightarrow 0$ ($\eta$ being already set to zero).

Actually, only the expression of~$\Psi'$ is given for this test-field solution:
\begin{eqnarray}
\label{eq_test_field}
\Psi'(R) = \frac{\pm q }{\left( 1 - \frac{R_{H}}{R} \right) \sqrt{\frac{4R}{R_{H}} - 3}},
\end{eqnarray}
where~$R_{H}$ is the Schwarzschild radius.
As stated earlier, this is sufficient because, due to the shift-symmetry of the theory, only the first derivatives of~$\phi$ are meaningful (and hence only~$\Psi'$ in static and spherical symmetry).

One can see from the action~(\ref{eq_action}) that flipping the sign of both~$\gamma$ and~$\phi$ of a given solution provides another solution to the theory.
This fact holds true in the limit~$\gamma \rightarrow 0$, which is why equation~(\ref{eq_test_field}) offers two test-field solutions with opposite signs.
Moreover, it is thus sufficient to seek solutions for positive~$\gamma$ only.

Once the Schwarzschild metric and the test scalar field~(\ref{eq_test_field}) are reexpressed in terms of the quasi-isotropic coordinates, the numerical process may converge to a solution of the system~(\ref{eq_metric_QI_N_bis})-(\ref{eq_scalar_bis}) in which the coupling~$\bar{\gamma}$ is set to a slightly nonzero value.
In turn, such solution serves as an initial guess to reach a solution with a slightly greater coupling~$\bar{\gamma}$ and so on.
% Referee 2: sole remark (part 3/3); Referee 1: remarks 3 and 4
Note that due to the Vainshtein mechanism (see section \ref{section_dynamics}) and the absence of kinetic term (see section \ref{section_BC}), no constraint can be inferred on the values of the parameters of the problem either from Solar System tests or cosmological arguments.
Therefore the values of $\bar{\gamma}$ picked for the graphs displayed in the present and later sections are only chosen so as to highlight with sufficient clarity how the results depend on $\bar{\gamma}$.
% End of "Referee 2: sole remark (part 3/3); Referee 1: remarks 3 and 4"

\begin{figure}
    \begin{center}
        \subfloat[Metric function~$N$]
        {
            \includegraphics[width=0.5\textwidth]{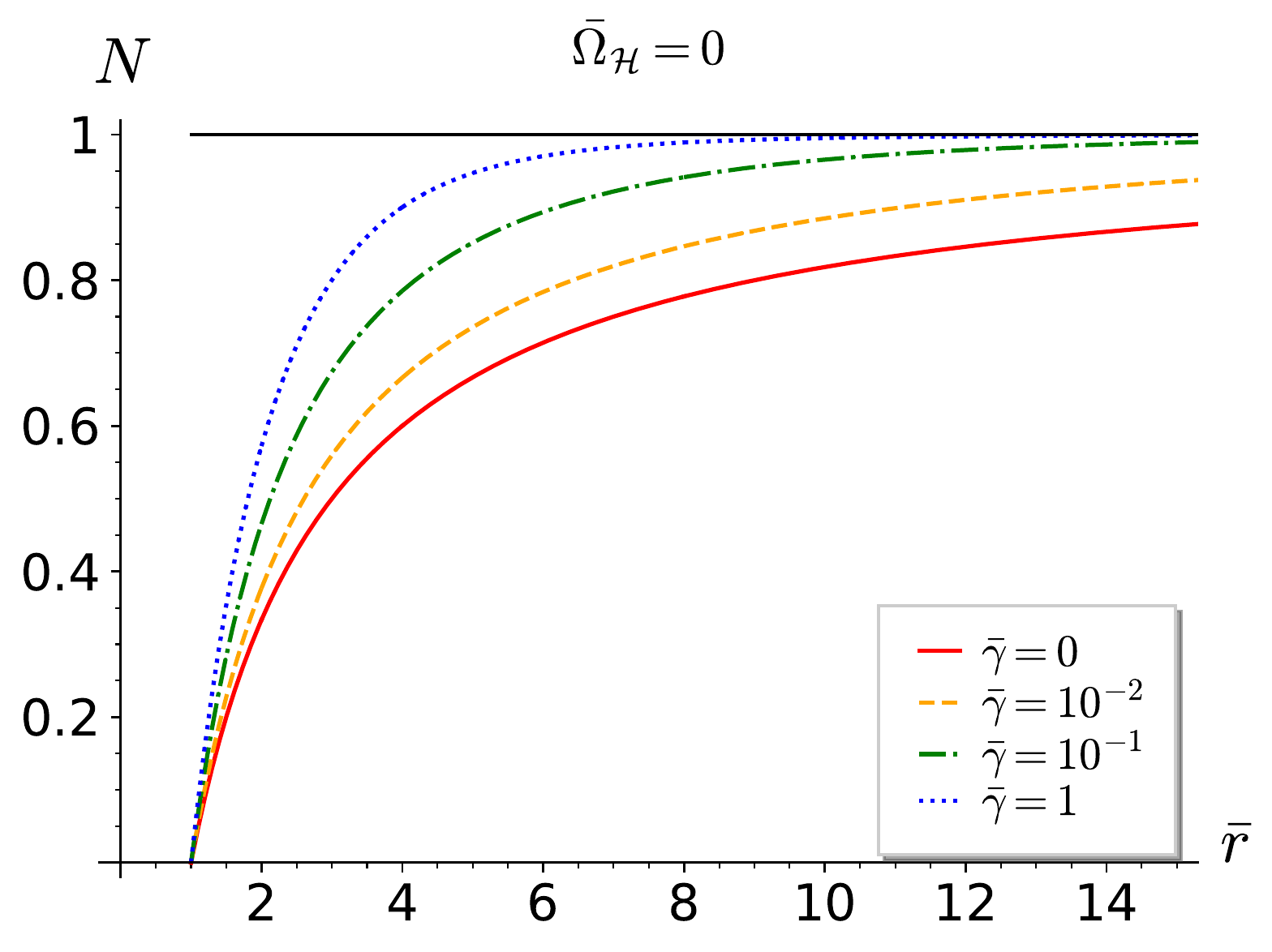}
            \label{fig_stat_sphe_N}
        }
        \subfloat[Metric function~$A$]
        {
            \includegraphics[width=0.5\textwidth]{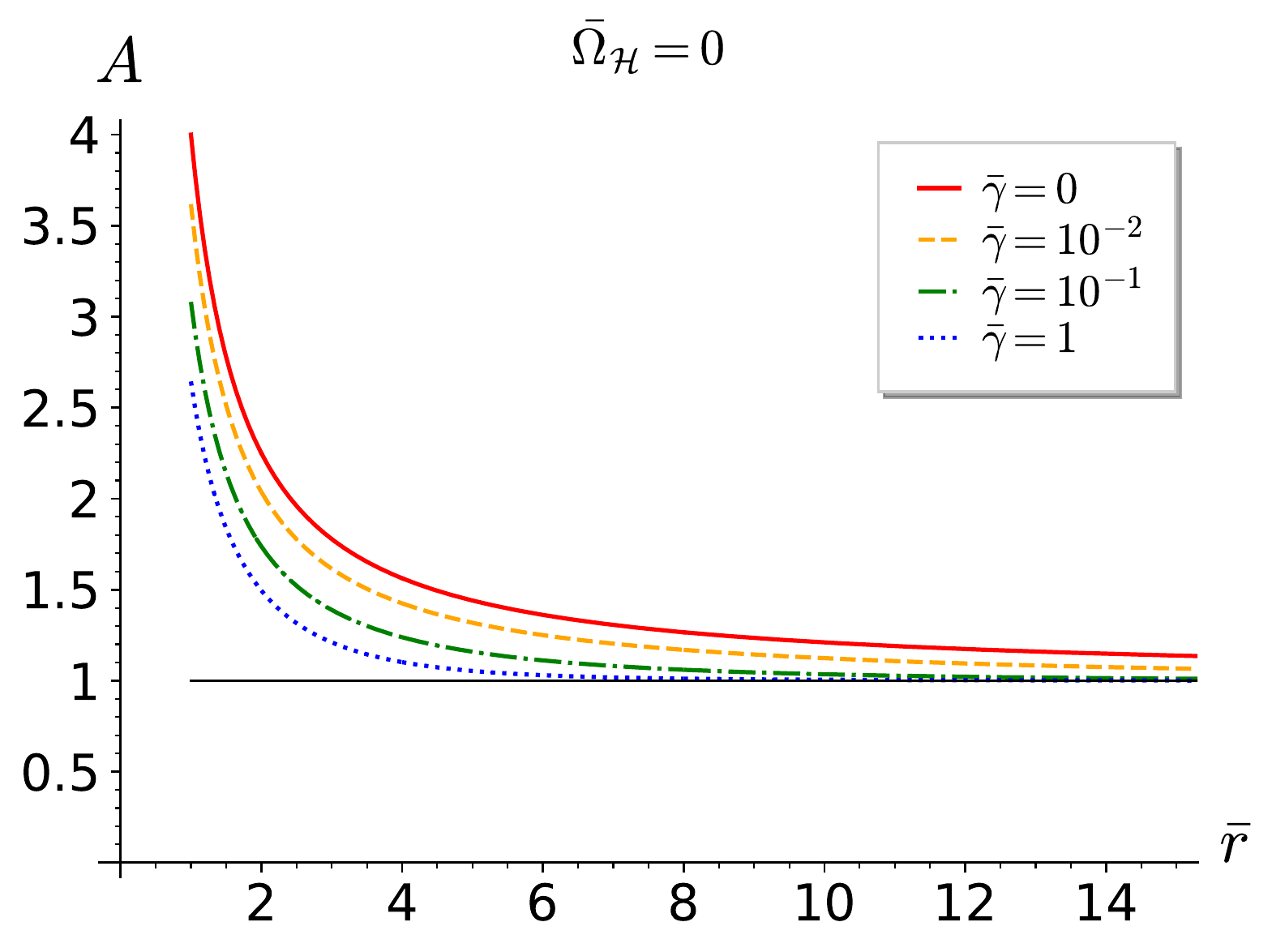}
            \label{fig_stat_sphe_A}
        } \\
        \subfloat[Regular scalar radial derivative~$Z \equiv N \bar{\Psi}'$]
        {
            \includegraphics[width=0.5\textwidth]{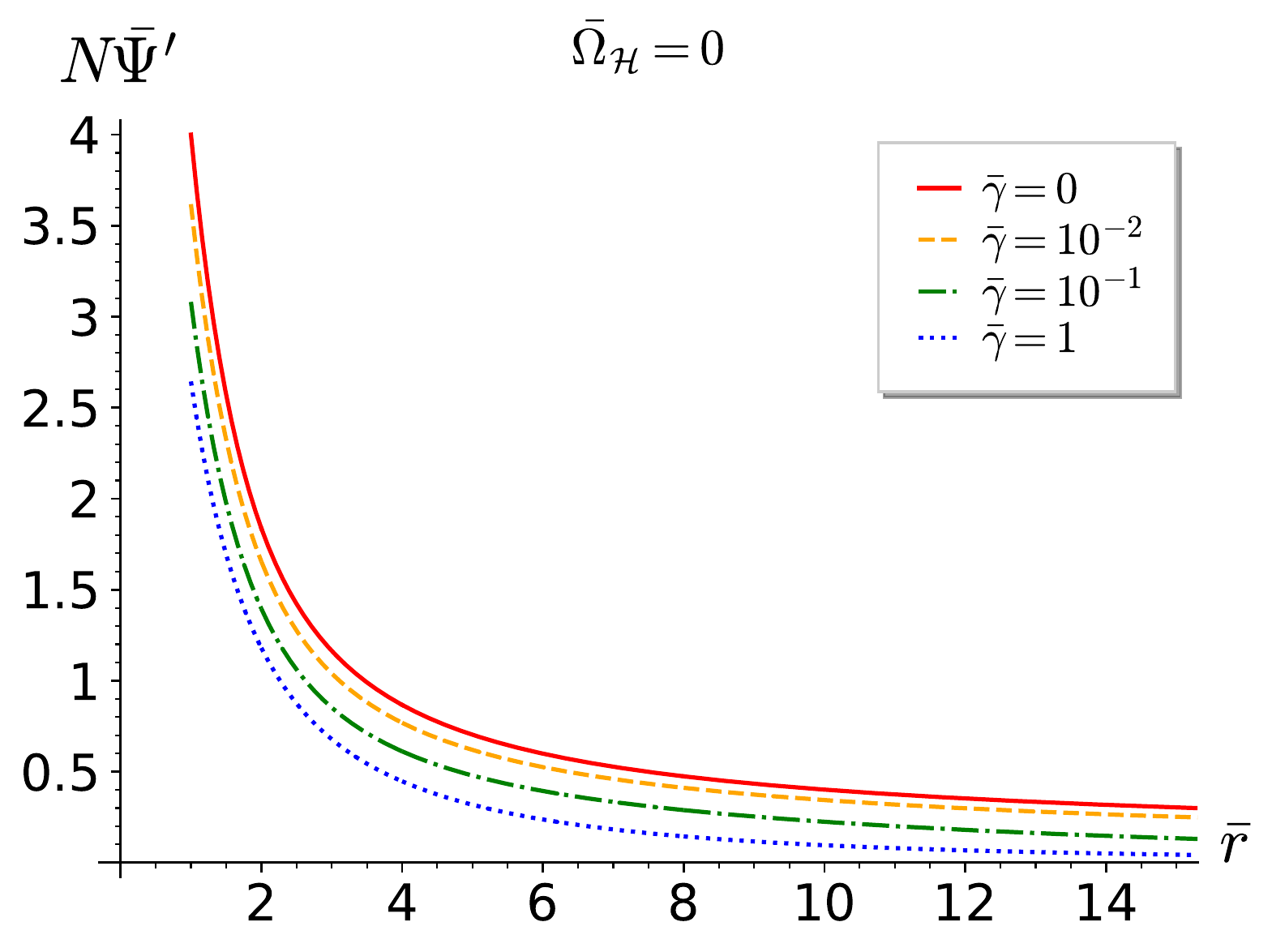}
            \label{fig_stat_sphe_Z}
        }
    \end{center}
\caption{Radial profiles in the static and spherically symmetric case ($\bar{\Omega}_{\mathcal{H}} = 0$) for values of~$\bar{\gamma}$ ranging from~$0$ to~$1$. When it is not zero, the limit at infinity is represented by a black, solid, horizontal asymptote.}
\label{fig_stat_sphe}
\end{figure}

The resulting solutions are displayed in Fig.~\ref{fig_stat_sphe}.
Due to spherical symmetry, one has~$B = A$, $\bar{\omega} = 0$ and~$\bar{\Psi}_{\theta} = 0$ everywhere, so that only the radial profiles of~$N$, $A$ and~$\bar{\Psi}'$ are non trivial.
Actually,~$Z \equiv N \bar{\Psi}'$ is plotted instead of~$\bar{\Psi}'$ because the former is finite on the horizon contrary to the latter.

For the function~$N$ (Fig.~\ref{fig_stat_sphe_N}), the boundary values~$N = 0$ at the horizon and~$N = 1$ at infinity are enforced according to section~\ref{section_BC}.
On the contrary, the values of~$A$ and~$Z$ on the horizon are not imposed due to the degeneracy of the equations.
Yet, it can be seen from the right-hand side~(\ref{eq_RHS_NA}) of equation~(\ref{eq_metric_QI_NA_bis}) that this degeneracy spontaneously imposes~$A^{2} = Z^{2}$ on the horizon, which is manifest on figures \ref{fig_stat_sphe_A} and \ref{fig_stat_sphe_Z} (and confirmed numerically).

One can also note that the greater the coupling~$\bar{\gamma}$ is, the faster the funtions~$N$, $A$ and~$Z$ converge towards their respective limits which correspond to a flat spacetime.
Then, when travelling from the horizon towards infinity, spacetime looks flat more rapidly for stronger coupling values~$\bar{\gamma}$.
In other terms, the more the scalar field back-reacts on the metric, the more it hides the deformations induced by the black hole.
This fact is further examined in section \ref{section_mass} below when discussing the extraction of a mass for these black hole solutions.

\subsection{Rotating black holes}
\label{section_rotating}

The Kerr metric is usually parametrized by  two parameters~$M$ (the mass) and~$a$ (the reduced angular momentum).
The radial coordinate~$r_{\mathcal{H}}$ of the event horizon may then be expressed in terms of these two parameters:
\begin{eqnarray}
\label{eq_horizon_QI_radius}
r_{\mathcal{H}} = \frac{M}{2} \sqrt{1 - \left(\frac{a}{M}\right)^{2}}.
\end{eqnarray}

Once~$r_{\mathcal{H}}$ is used to make all the quantities dimensionless and all the metric components are expressed in terms of~$\bar{r} \equiv r/r_{\mathcal{H}}$, the dimensionless Kerr solution is only parametrized by one quantity, which can be chosen to be~$\bar{\Omega}_{\mathcal{H}}$.
Of course, one such quantity might not be enough to parametrize the whole set of black hole solutions of the cubic Galileon theory with a scalar field structured as~(\ref{eq_scalar_ansatz}).
Yet, the numerical approach employed here only reaches the solutions that continuously connect to Schwarzschild, by increasing~$\bar{\gamma}$ first and then~$\bar{\Omega}_{\mathcal{H}}$.
This is why, once~$\bar{\gamma}$ is fixed,~$\bar{\Omega}_{\mathcal{H}}$ is also the only quantity that parametrizes the solutions presented here.

\begin{figure}
    \begin{center}
        \subfloat[Metric function~$N$ at~$\theta = \pi/2$]
        {
            \includegraphics[clip, trim=0cm 0cm 4cm 2cm, width=0.5\textwidth]{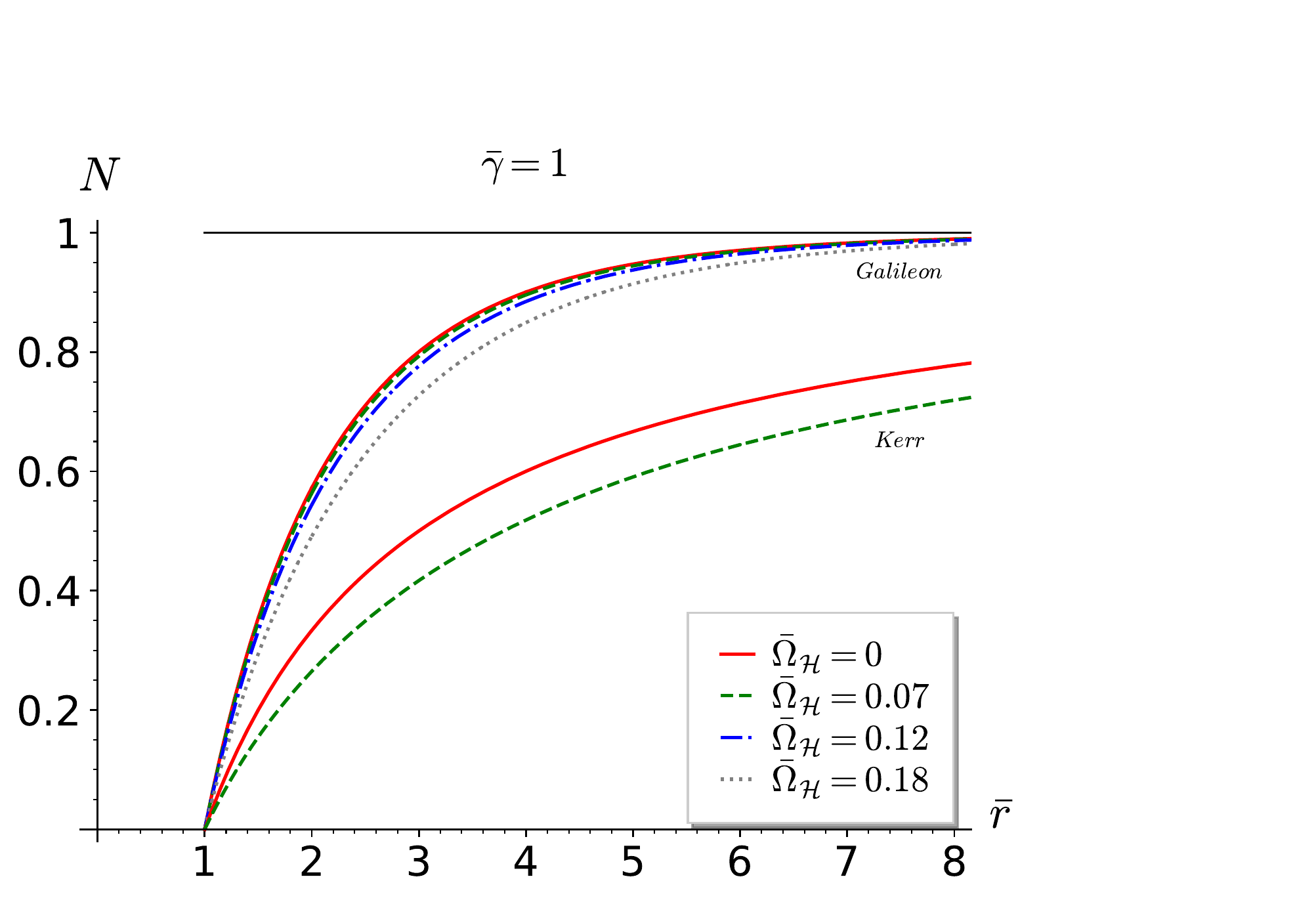}
            \label{fig_rot_N}
        }
        \subfloat[Metric function~$A$ at~$\theta = \pi/2$]
        {
            \includegraphics[clip, trim=0cm 0cm 0cm 2cm, width=0.5\textwidth]{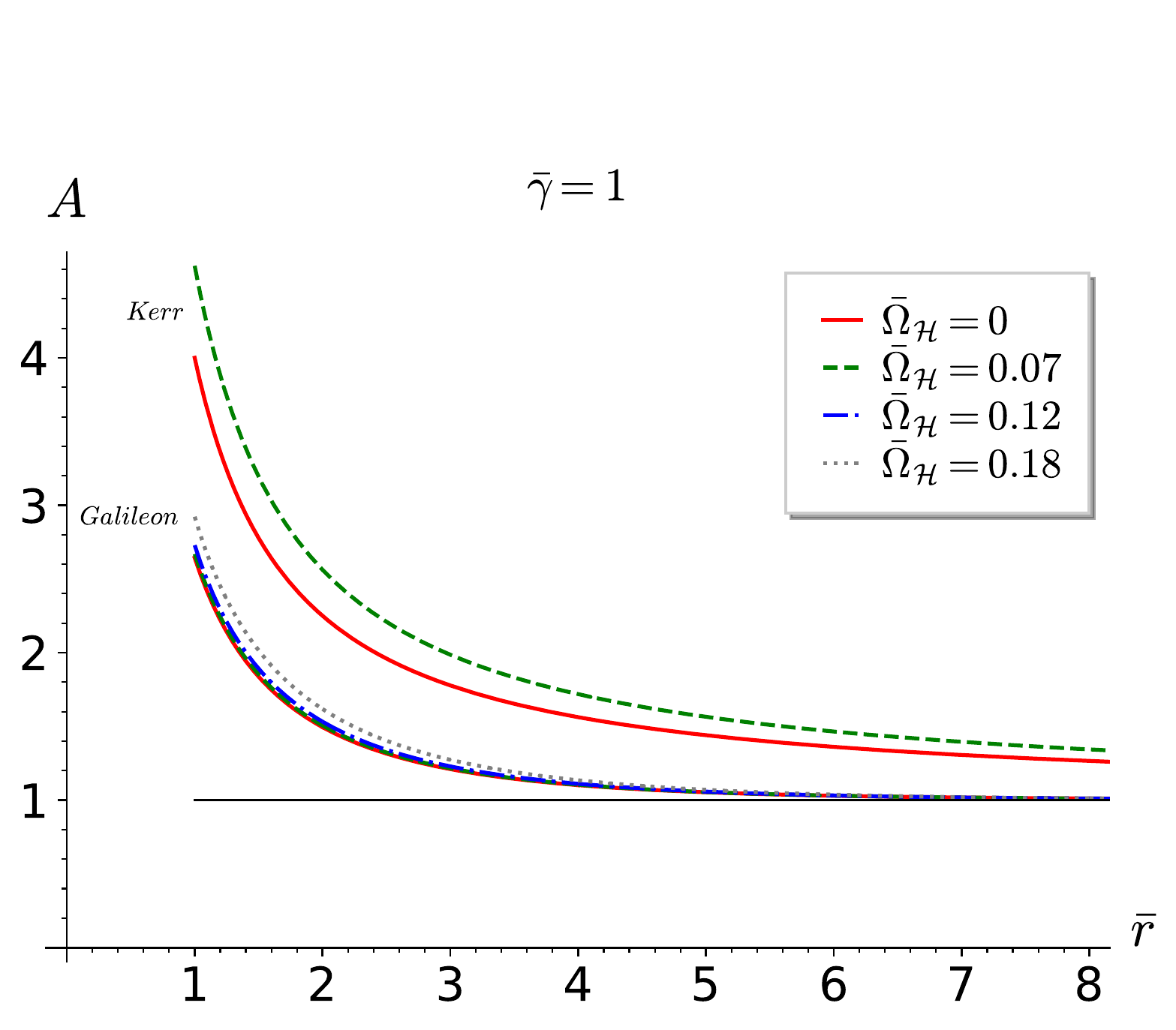}
            \label{fig_rot_A}
        } \\
        \subfloat[Metric function~$B$ at~$\theta = \pi/2$]
        {
            \includegraphics[width=0.5\textwidth]{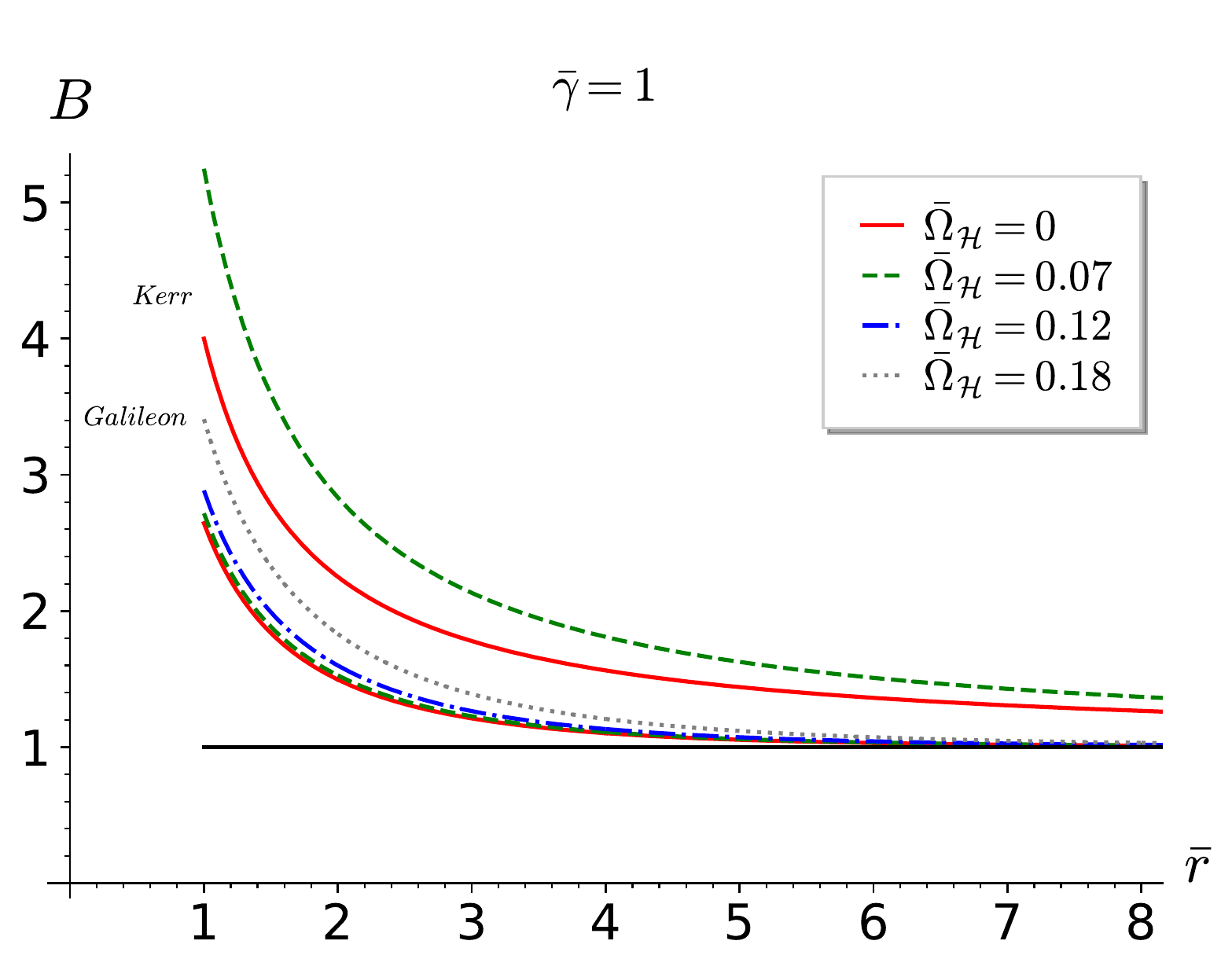}
            \label{fig_rot_B}
        }
        \subfloat[Metric function~$\bar{\omega}$ at~$\theta = \pi/2$]
        {
            \includegraphics[width=0.5\textwidth]{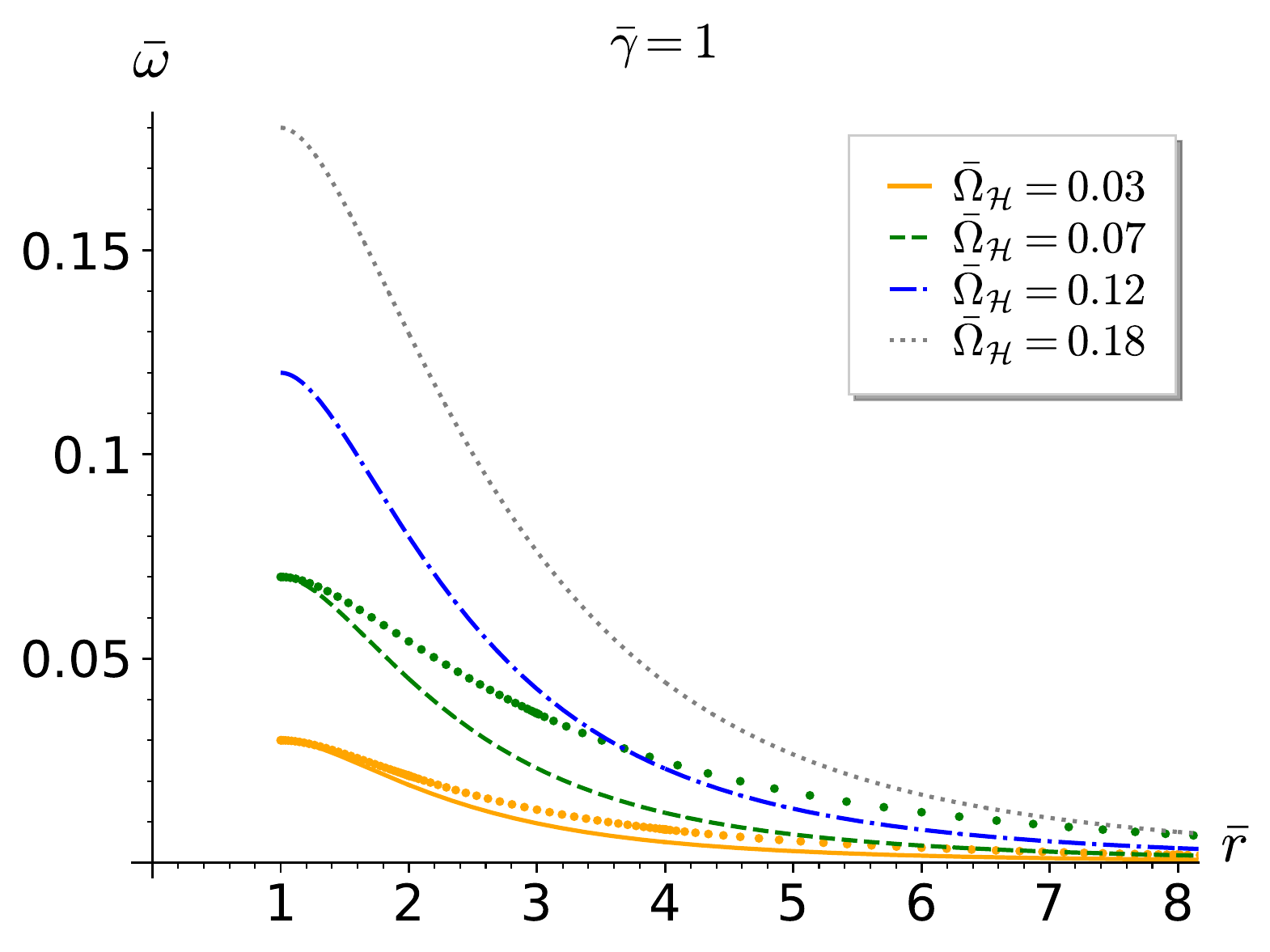}
            \label{fig_rot_adOm}
        } \\
        \subfloat[Regular scalar radial derivative~$Z \equiv N \bar{\Psi}'$ at $\theta = \pi/2$]
        {
            \includegraphics[width=0.5\textwidth]{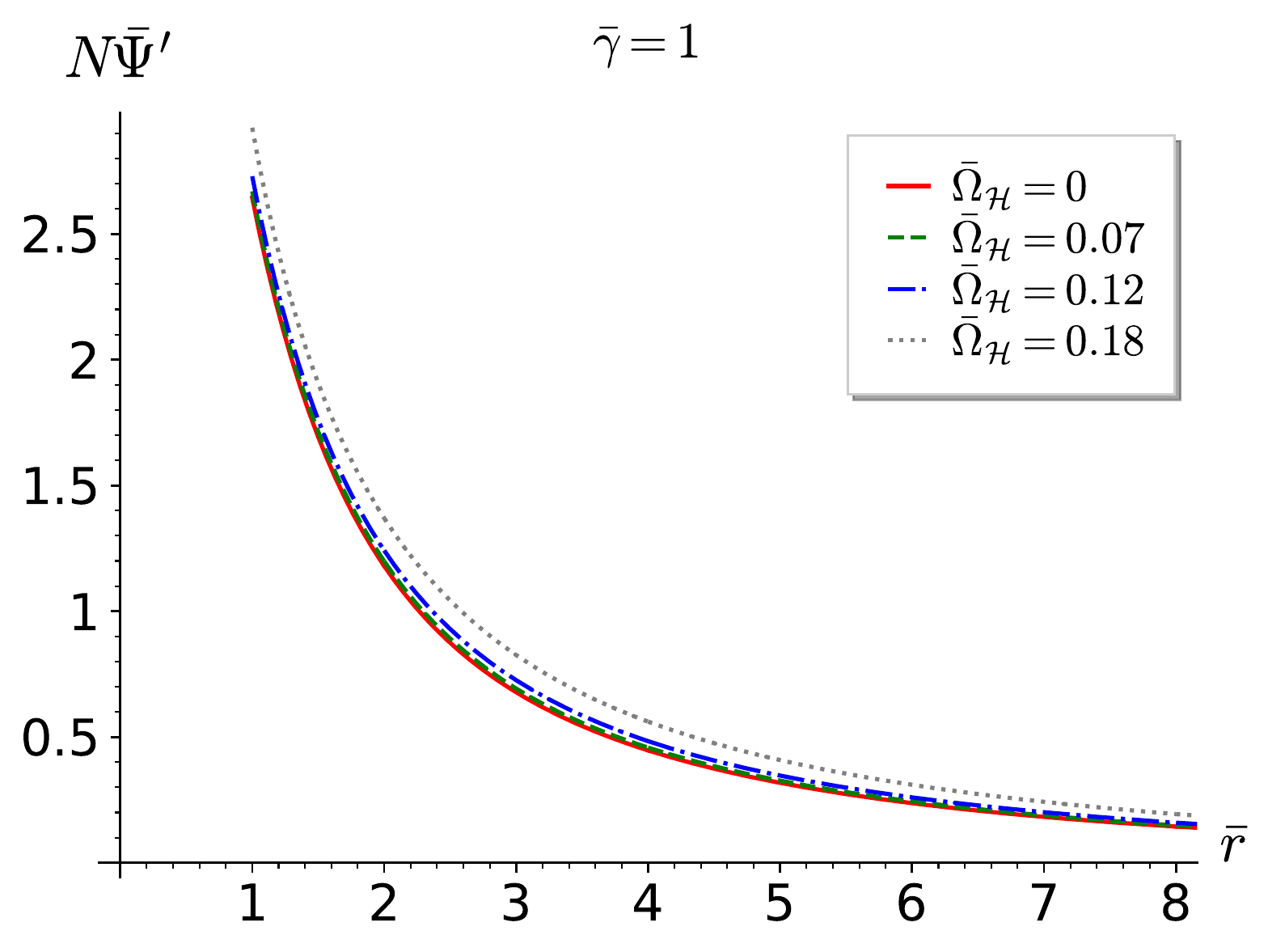}
            \label{fig_rot_Z}
        }
        \subfloat[Scalar angular derivative~$\bar{\Psi}_{\theta}$ at~$\theta = \pi/4$]
        {
            \includegraphics[width=0.5\textwidth]{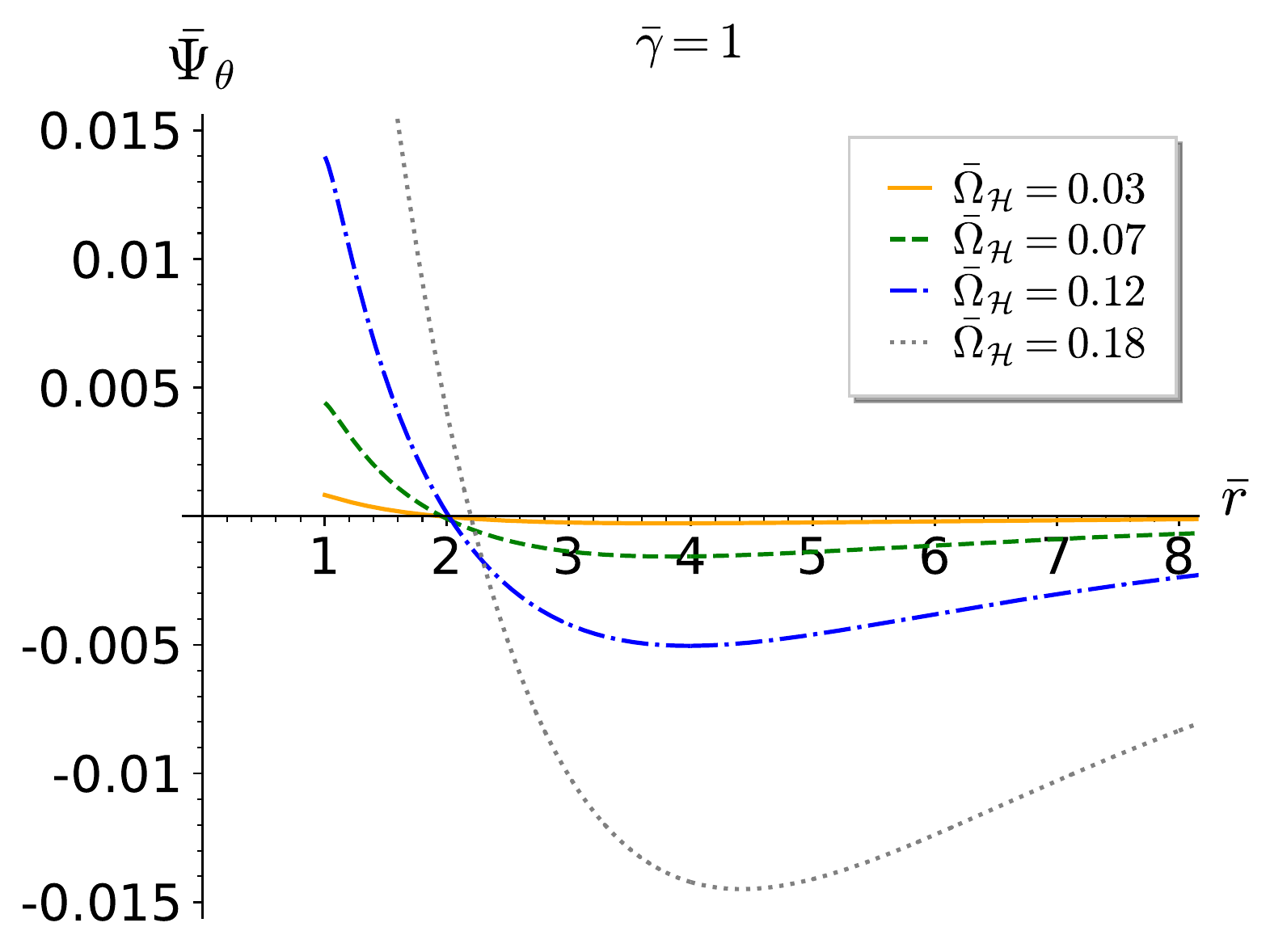}
            \label{fig_rot_dthPsi}
        }
    \end{center}
\caption{Radial profiles at fixed coupling~$\bar{\gamma} = 1$ and different~$\bar{\Omega}_{\mathcal{H}}$. When it is not zero, the limit at infinity is represented by a black, solid, horizontal asymptote.}
\label{fig_rot}
\end{figure}

Figure~\ref{fig_rot} displays the radial profiles of all six functions~$N$, $A$, $B$, $\bar{\omega}$, $\bar{\Psi}'$ and~$\bar{\Psi}_{\theta}$ at fixed~$\bar{\gamma} = 1$ for various values of~$\bar{\Omega}_{\mathcal{H}}$.
For~$\bar{\Omega}_{\mathcal{H}} = 0$ and~$0.07$, the corresponding dimensionless Kerr solution is plotted for comparison: in the case of~$N$, $A$ and~$B$, the Kerr curve has the same color and linestyle as the Galileon curve corresponding to the same parameter~$\bar{\Omega}_{\mathcal{H}}$, and in the case of~$\bar{\omega}$, it is the thick dotted curve having the same value at the horizon with its Galileon analog.
As for~$Z$ and~$\bar{\Psi}_{\theta}$, no Kerr analog is displayed since no test-field solutions are known in the rotating case (i.e. solutions to the scalar equation~(\ref{eq_scalar}) on a Kerr background with the scalar ansatz~(\ref{eq_scalar_ansatz})) and such solutions could not be obtained numerically.

\begin{figure}
    \begin{center}
    \includegraphics[width=0.5\textwidth]{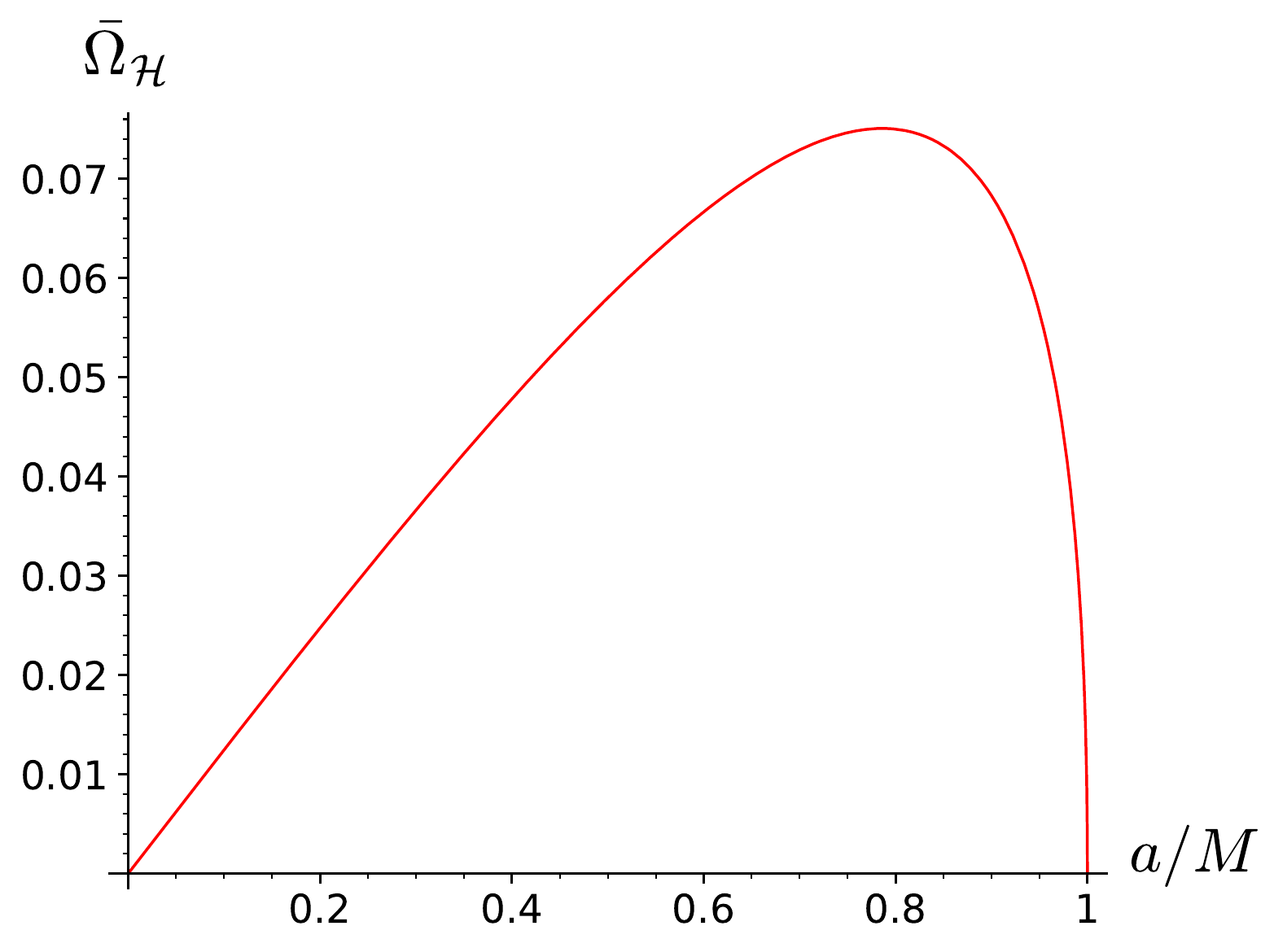}
    \end{center}
\caption{$\bar{\Omega}_{\mathcal{H}}$ with respect to~$a/M$ for Kerr black holes.}
\label{figure_adOm_vs_asM}
\end{figure}

For the other values of~$\bar{\Omega}_{\mathcal{H}}$ ($0.12$ and~$0.18$), the Galileon solutions displayed in figure~\ref{fig_rot} do not admit a Kerr analog.
The reason is that the cubic Galileon admits solutions with dimensionless angular velocities greater than the maximum~$\bar{\Omega}_{\mathcal{H}}$ that can be obtained from the Kerr metric.
More precisely, at fixed mass~$M$, the angular velocity~$\Omega_{\mathcal{H}}$ of the Kerr black hole cancels at~$a/M = 0$ and monotonically increases towards a finite value at~$a/M = 1$, while the radial quasi-isotropic coordinate~$r_{\mathcal{H}}$ of the event horizon is finite at~$a/M = 0$ and monotonically decreases towards~$0$ at~$a/M = 1$ according to equation (\ref{eq_horizon_QI_radius}).
Then,~$\bar{\Omega}_{\mathcal{H}} = r_{\mathcal{H}} \Omega_{\mathcal{H}}$ is a positive function of the dimensionless ratio~$a/M \in [0,1]$ cancelling both at~$0$ and~$1$:
\begin{eqnarray}
\bar{\Omega}_{\mathcal{H}} = \frac{1}{4} \frac{ \frac{a}{M} }{1 + \left(1 - \left(\frac{a}{M}\right)^{2}\right)^{-\frac{1}{2}}},
\end{eqnarray}
which is plotted on Fig.~\ref{figure_adOm_vs_asM}.
In particular, this function has a maximum value~$\bar{\Omega}_{\mathcal{H},\rm max} \simeq 0.075$ at~$a/M \simeq 0.8$, which actually turns out to be possible to exceed in the cubic Galileon theory.
This will appear clearly in section \ref{section_phys} when extracting the angular momentum and the surface gravity of these black hole solutions.

\begin{figure}
    \begin{center}
    \includegraphics[width=0.5\textwidth]{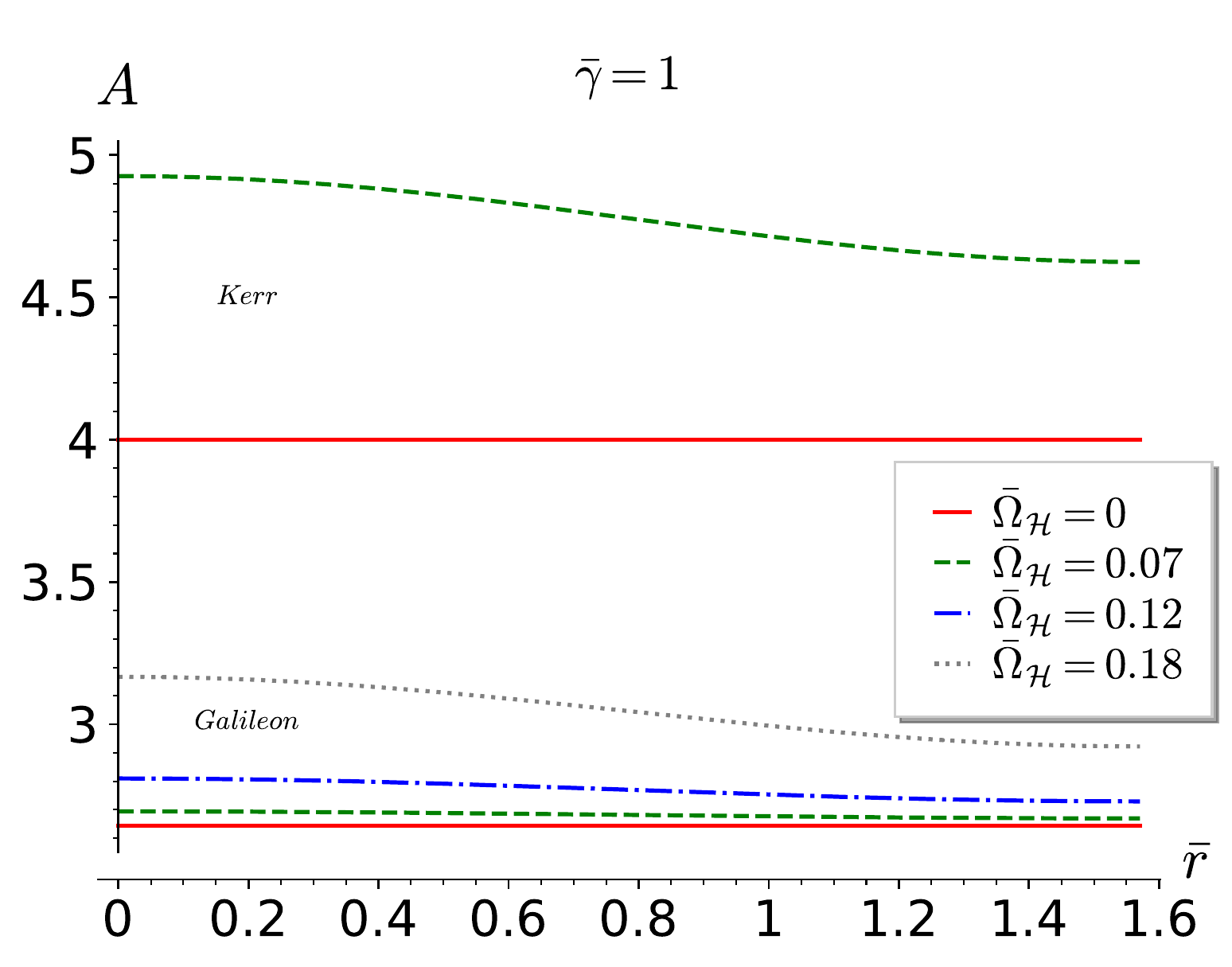}
    \end{center}
\caption{Angular profile of~$A$ at the horizon.}
\label{fig_rot_A_angular_horizon}
\end{figure}

Going back to figure~\ref{fig_rot}, one first notes that, although the global behaviours are the same, there are non negligible gaps near the horizon between the Galileon solution and Kerr for any fixed dimensionless angular velocity.
Naturally, for both the Galileon and Kerr, increasing~$\bar{\Omega}_{\mathcal{H}}$ tends to slow the convergence towards the asymptotic values (at fixed radial coordinate, it is expected that spacetime looks less flat if the hole is rotating).
One last remark to make is that, although these solutions feature quite rapid rotation ($\bar{\Omega}_{\mathcal{H}} = 0.07$ corresponds to~$a/M \simeq 0.65$ for Kerr), the angular variations of the various functions are quite moderate for both the Galileon and Kerr; this is manifest on Fig.~\ref{fig_rot_A_angular_horizon} which displays the angular profile of the function~$A$ on the horizon.

\section{Physical properties}
\label{section_phys}

    \subsection{Mass}
    \label{section_mass}

The general definition of the Komar mass of an asymptotically flat stationary spacetime, equipped with a foliation~$\left( \Sigma_{t} \right)_{t \in \mathbb{R}}$ by spacelike hypersurfaces, is \cite{3p1_formalism,intro_relat_stars}
\begin{eqnarray}
\label{eq_Komar_mass_general}
M_{\rm Komar} \equiv - \frac{1}{8\pi} \int_{\mathcal{S}} \ast d\xi,
\end{eqnarray}
where~$\xi$ is the stationary Killing vector here identified with its metric dual form,~$\ast$ is the Hodge star and~$\mathcal{S} \subset \Sigma_{t_{0}}$ (for some~$t_{0} \in \mathbb{R}$) is a closed spacelike 2-surface containing the intersection of~$\Sigma_{t_{0}}$ with the support of the energy-momentum tensor.
In general relativity, the Einstein equations guarantee that the Komar mass does not depend on the choice of such 2-surface~$\mathcal{S}$.

In practice, one then usually uses a 2-surface~$\mathcal{S}$ lying at spatial infinity.
In particular, in quasi-isotropic coordinates, the Komar mass may be computed from the following integral:
\begin{eqnarray}
\label{eq_Komar_mass_QI}
M_{\rm Komar} = \frac{1}{2} \lim_{r \rightarrow \infty} \int_{0}^{\pi} \partial_{r}N\ r^{2} \sin\theta d\theta.
\end{eqnarray}

Therefore, if~$N$ has the following asymptotic behaviour:
\begin{eqnarray}
\label{eq_N_asympt}
N = 1 + \frac{N_{1}}{r} + o\left( \frac{1}{r} \right),
\end{eqnarray}
where~$N_{1}$ is a constant, then
\begin{eqnarray}
\label{eq_Komar_mass_N}
M_{\rm Komar} = - N_{1}.
\end{eqnarray}

In the cubic Galileon theory, the contribution from the scalar field into equation~(\ref{eq_metric}) does not allow to guarantee that the expression~(\ref{eq_Komar_mass_general}) is independent of the 2-surface~$\mathcal{S}$.
Yet, as is usually done, one may try to extract a mass from the relation~(\ref{eq_Komar_mass_N}).
This can be done explicitly in the static and spherically symmetric case.

To do so, it is simpler to first switch back to the Schwarzschild-like coordinates~(\ref{eq_metric_Schw_like}) used in section \ref{section_BC} to extract the asymptotic behaviours~(\ref{eq_asympt_h})-(\ref{eq_asympt_chi}) when~$\eta \neq 0$.
Repeating the same procedure in the case of asymptotic flatness, i.e. injecting expansions in~$1/R$ into~(\ref{eq_4_1})-(\ref{eq_4_3}) with~$\eta = \Lambda = 0$, one finds the following asymptotic behaviours:
\begin{eqnarray}
\label{eq_asympt_chi_flat}
h(R) \Psi'(R) = \frac{d}{R^{2}} + O\left( \frac{1}{R^{5}} \right), \\
\label{eq_asympt_h_flat}
h(R) = 1 - \frac{d^{2}}{q^{2} R^{4}} + O\left( \frac{1}{R^{7}} \right), \\
\label{eq_asympt_f_flat}
f(R) = 1 - \frac{4 d^{2}}{q^{2} R^{4}} + O\left( \frac{1}{R^{7}} \right),
\end{eqnarray}
where~$d$ is some fixed constant.
Note here that the test field approximation~(\ref{eq_test_field}) gives a wrong indication about the asymptotic behavior of~$\Psi'(R)$ since it behaves as~$1/\sqrt{R}$ although, according to equation~(\ref{eq_asympt_chi_flat}), it behaves as~$1/R^{2}$ as soon as the coupling~$\gamma$ is nonzero, no matter how small.
Yet this did not prevent the test-field solution from being useful as an initial guess in the numerical procedure.

Now, the change of coordinates from the Schwarzschild-like coordinates~$(t,R,\theta,\varphi)$ to the quasi-isotropic coordinates~$(t,r,\theta,\varphi)$ is merely given by the positive function~$R(r)$ defined on~$[r_{\mathcal{H}}, +\infty)$ such that
\begin{eqnarray}
\label{eq_change_Schw_QI}
r R'(r) = R(r) \sqrt{f\left( R(r) \right)}.
\end{eqnarray}

From this, one can infer the same types of asymptotic behaviours as~(\ref{eq_asympt_chi_flat})-(\ref{eq_asympt_f_flat}) for the functions~$Z$, $N$ and~$A$:
\begin{eqnarray}
\label{eq_asympt_Z_flat}
Z(r) = \frac{e}{r^{2}} + o\left( \frac{1}{r^{2}} \right), \\
\label{eq_asympt_N_flat}
N(r) = 1 + \frac{e'}{r^{4}} + o\left( \frac{1}{r^{4}} \right), \\
\label{eq_asympt_A_flat}
A(r) = 1 + \frac{e''}{r^{4}} + o\left( \frac{1}{r^{4}} \right),
\end{eqnarray}
where~$e$, $e'$ and~$e''$ are some fixed constants.

\begin{figure}
    \begin{center}
        \subfloat[log-log graph of~$1-N$.]
        {
            \includegraphics[width=0.5\textwidth]{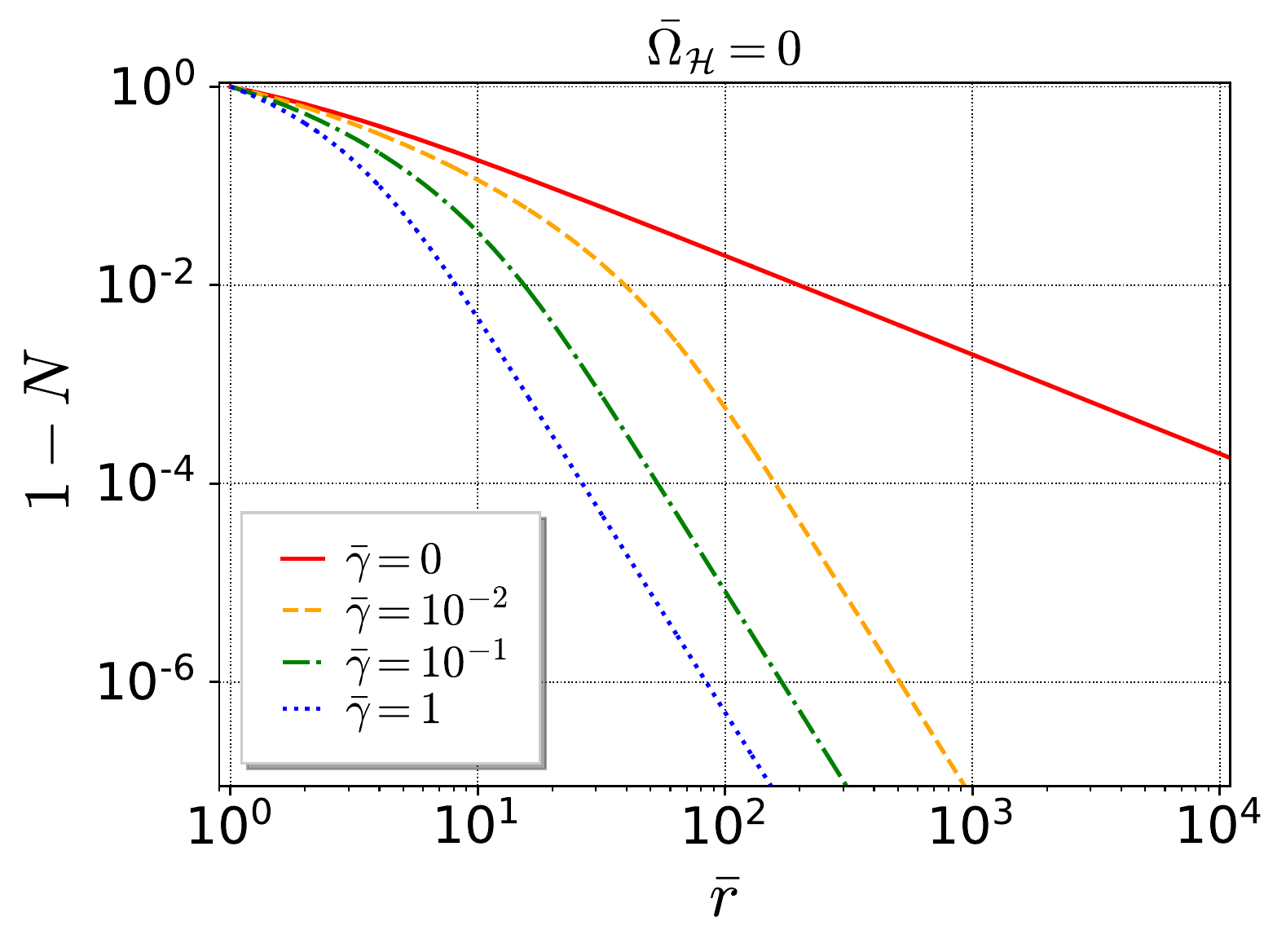}
            \label{fig_stat_1-N_loglog}
        }
        \subfloat[log-log graph of~$Z \equiv N \bar{\Psi}'$.]
        {
            \includegraphics[width=0.5\textwidth]{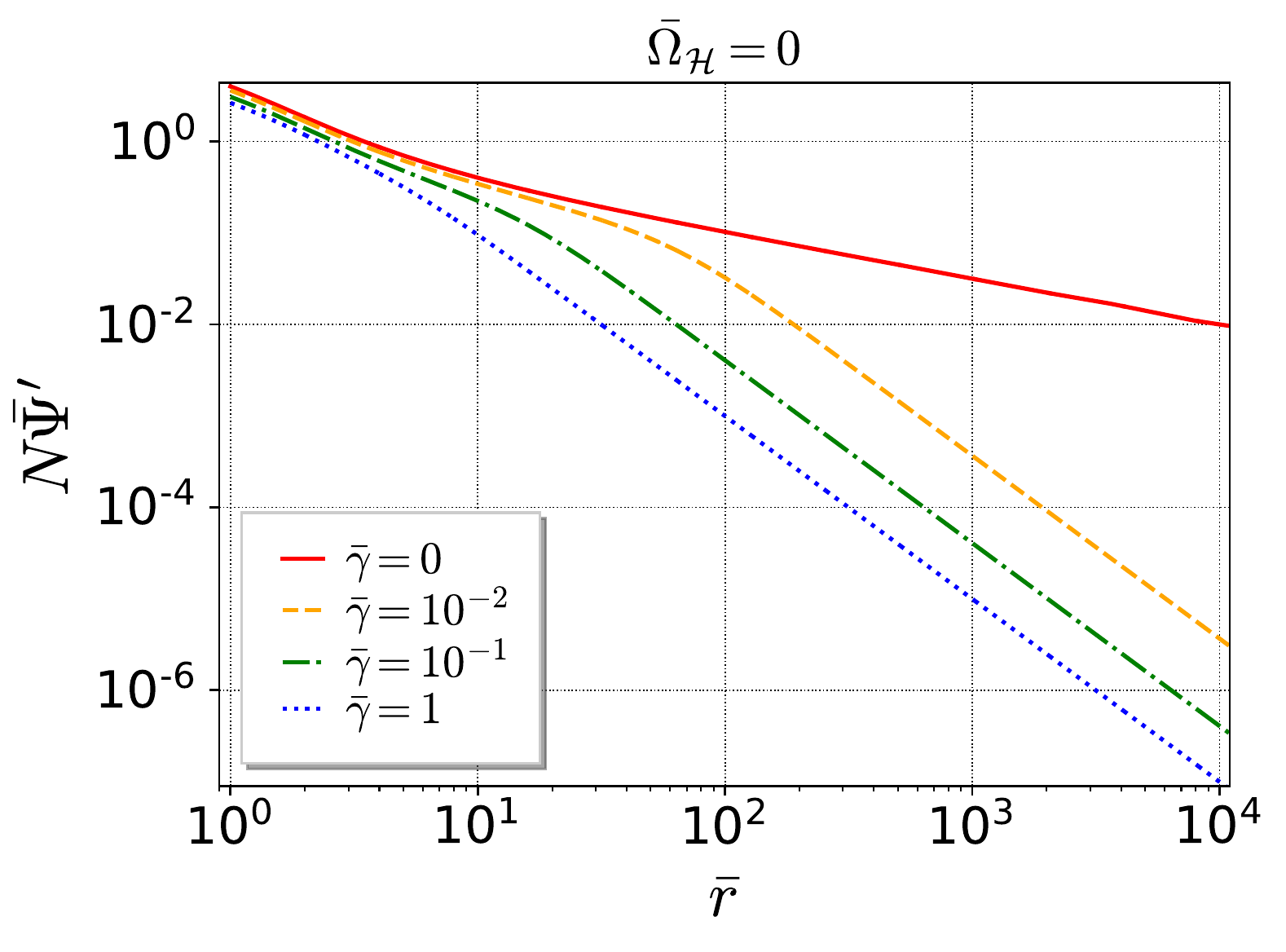}
            \label{fig_stat_Z_loglog}
        }
    \end{center}
\caption{Asymptotic behaviours in the static and spherically symmetric case.}
\label{fig_stat_loglog}
\end{figure}

One concludes that there is no term to the first inverse power of~$r$ in the expansion~(\ref{eq_asympt_N_flat}) of~$N$, meaning that the Komar mass is zero according to the relation~(\ref{eq_Komar_mass_N}).
This fact may be checked numerically by extracting the asymptotic slope of~$1-N$ in a log-log graph (Fig.~\ref{fig_stat_1-N_loglog}), which corresponds to the asymptotically dominant power of~$r$; the resulting numerical value is perfectly consistent with~$-4$.
The function~$A-1$ does have a very similar log-log graph, and one may check on Fig.~\ref{fig_stat_Z_loglog} that, for the function~$Z$, the asymptotic slope is numerically consistent with~$-2$.

Such asymptotic behaviours seem to be maintained in the rotating case although the dominant power for~$N$ might not be exactly~$-4$, but still smaller than~$-3.5$, hence no mass term can be extracted either.
One is thus led to conclude that the presence of a scalar field with structure~(\ref{eq_scalar_ansatz}) in the cubic Galileon theory generically hides the mass of an asymptotically flat black hole from infinity.
Note that this could not be the case whenever asymptotic flatness is abandoned, i.e. nonzero~$\Lambda$ and/or~$\eta$, since the asymptotic expansions (4.17) of \cite{cubic_BH} require a standard mass term from the first inverse power of~$r$.

    \subsection{Angular momentum}
    \label{section_angular_momentum}

Similarly to the definition~(\ref{eq_Komar_mass_general}), the Komar angular momentum of an asymptotically flat axisymmetric spacetime is defined as
\begin{eqnarray}
\label{eq_Komar_angul_general}
J_{\rm Komar} \equiv \frac{1}{16\pi} \int_{\mathcal{S}} \ast d\chi,
\end{eqnarray}
where~$\xi$ is the axisymmetric Killing vector.

Using the quasi-isotropic coordinates, the definition~(\ref{eq_Komar_angul_general}) reexpresses as
\begin{eqnarray}
\label{eq_Komar_angul_QI}
J_{\rm Komar} = - \frac{1}{8} \lim_{r \rightarrow \infty} \int_{0}^{\pi} \partial_{r}\omega\ r^{4} \sin^{3}\theta d\theta.
\end{eqnarray}

Therefore, if~$\omega$ has the following asymptotic behaviour:
\begin{eqnarray}
\label{eq_omega_asympt}
\omega = \frac{\omega_{1}}{r^{3}} + o\left( \frac{1}{r^{3}} \right),
\end{eqnarray}
where~$\omega_{1}$ is a constant, then
\begin{eqnarray}
\label{eq_Komar_omega}
J_{\rm Komar} = \frac{\omega_{1}}{2}.
\end{eqnarray}

\begin{figure}
    \begin{center}
        \subfloat[log-log graph of function~$\bar{\omega}$.]
        {
            \includegraphics[width=0.5\textwidth]{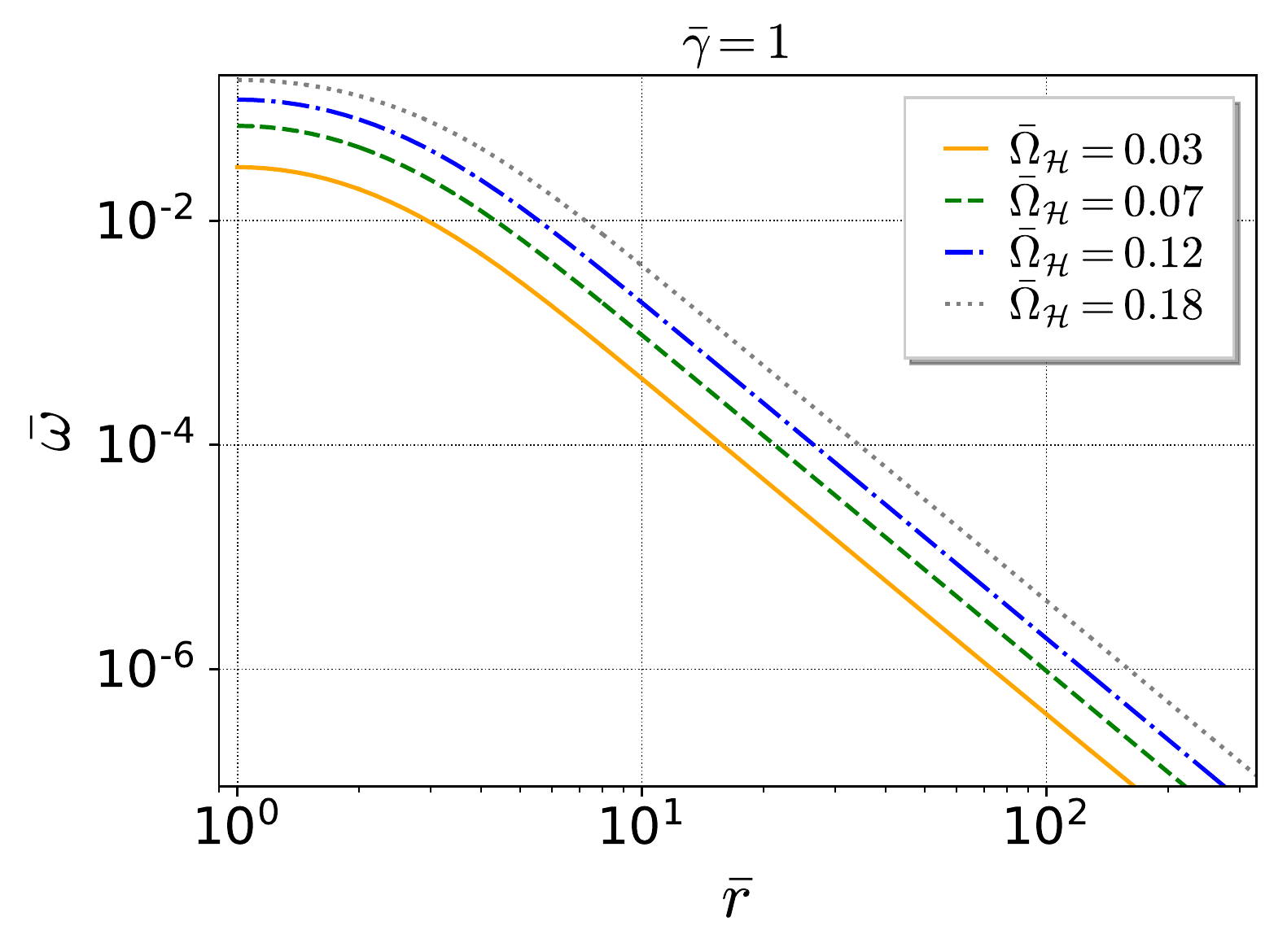}
            \label{fig_rot_adomega_loglog}
        }
        \subfloat[Angular velocity with respect to angular momentum.]
        {
            \includegraphics[width=0.5\textwidth]{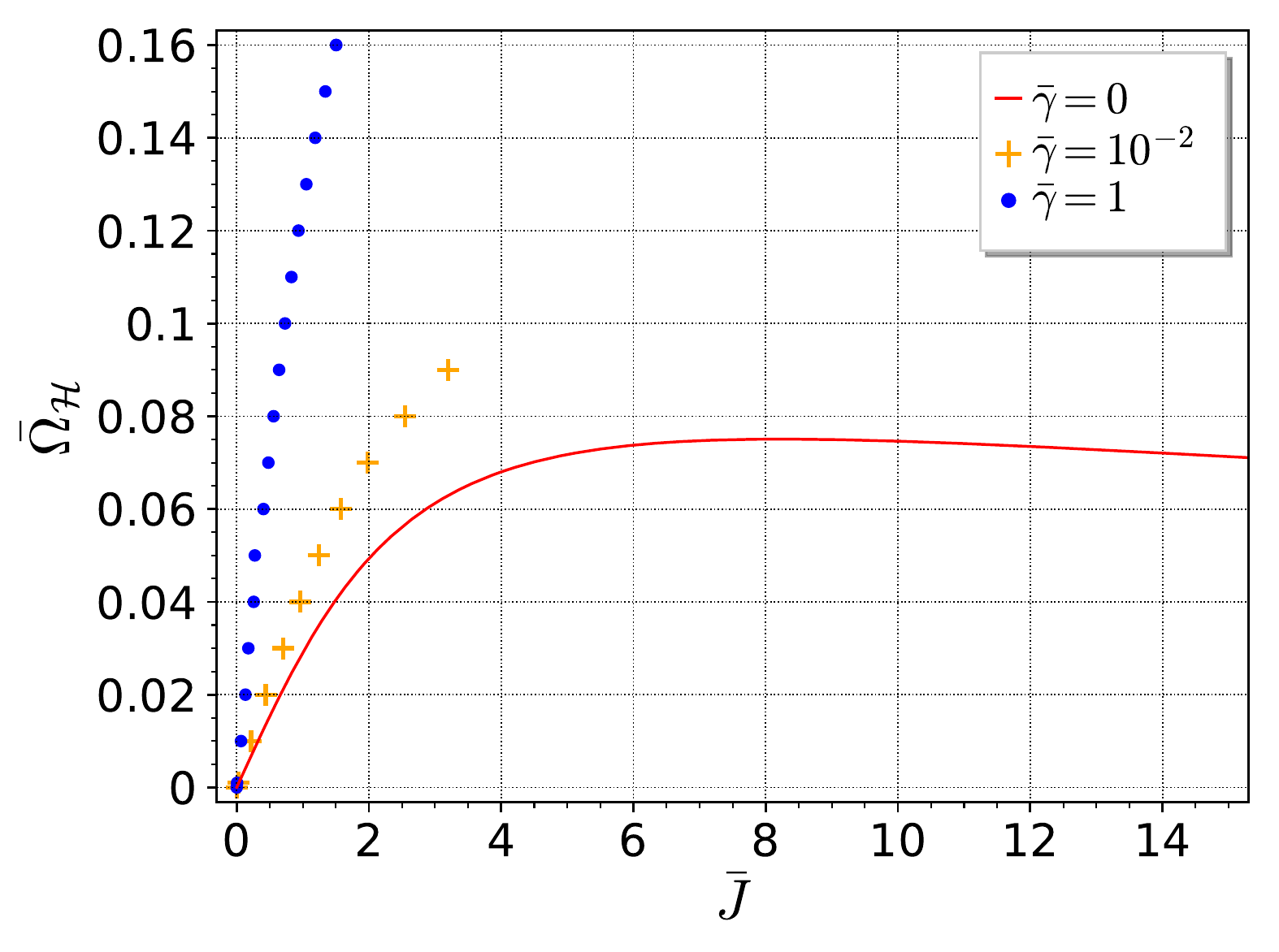}
            \label{fig_adOm_vs_adJ}
        }
    \end{center}
\caption{Angular momentum extracted from the asymptotic behaviour of~$\bar{\omega}$.}
\label{fig_rot_loglog_adJ}
\end{figure}

Again, one may try to extract a Komar angular momentum from the asymptotic expansion of~$\omega$ although, in the cubic Galileon theory, such a value would have no reason to be common to all other 2-surfaces~$\mathcal{S}$.
Figure~\ref{fig_rot_adomega_loglog} confirms that~$\bar{\omega}$ has the asymptotic behaviour (\ref{eq_omega_asympt}) (asymptotic slope equal to~$-3$) so that the Komar angular momentum is nonzero.

Since only dimensionless quantities are processed numerically, one has
\begin{eqnarray}
\label{eq_adomega_asympt}
\bar{\omega} \equiv r_{\mathcal{H}} \omega \sim \frac{2 \bar{J}_{\rm Komar}}{\bar{r}^{3}},
\end{eqnarray}
where~$\bar{J}$ is the dimensionless Komar angular momentum:
\begin{eqnarray}
\label{eq_adJ}
\bar{J}_{\rm Komar} = \frac{J_{\rm Komar}}{r_{\mathcal{H}}^{2}}.
\end{eqnarray}

The values of~$\bar{J}_{\rm Komar}$ extracted for all the~$\bar{\Omega}_{\mathcal{H}}$ that were reached for~$\bar{\gamma} = 10^{-2}$ and~$1$ are marked in figure~\ref{fig_adOm_vs_adJ}.
The relation between~$\bar{\Omega}_{\mathcal{H}}$ and~$\bar{J}_{\rm Komar}$ can be expressed explicitly in the case of the Kerr family, and it is represented by the solid red curve to highlight the deviations from GR.

As mentioned in section \ref{section_rotating},~$r_{\mathcal{H}}$ tends to zero for the extremal Kerr solutions while~$J_{\rm Komar}$ tends to the finite value~$M^{2}$.
Therefore~$\bar{\Omega}_{\mathcal{H}}$ goes to zero while~$\bar{J}_{\rm Komar}$ diverges according to the relation (\ref{eq_adJ}).
This is why the curve corresponding to Kerr in Fig.~\ref{fig_adOm_vs_adJ} is defined all over~$\mathbb{R}_{+}$ and converges to zero at infinity.
Since~$\bar{\Omega}_{\mathcal{H}} = 0$ for~$\bar{J}_{\rm Komar} = 0$ and~$\bar{\Omega}_{\mathcal{H}}$ is positive, it must also have a maximum which is reached for~$\bar{J}_{\rm Komar} \simeq 8$ according to Fig.~\ref{fig_adOm_vs_adJ}.
One can see that some cubic Galileon solutions exceed this maximum value, which clearly shows why it was not possible to provide a Kerr analog for the metric functions in Fig.~\ref{fig_rot} for~$\bar{\gamma} = 0.12$ and~$0.18$.

Yet the existence of a maximum value for~$\bar{\Omega}_{\mathcal{H}}$ in the Kerr case reveals that this quantity does not provide a bijective parametrization of the families of dimensionless black hole solutions.
This represents a numerical difficulty: the solutions are gradually constructed by increasing the parameter~$\bar{\Omega}_{\mathcal{H}}$ starting from the static and spherically symmetric solution~$(\bar{\Omega}_{\mathcal{H}},\bar{J}_{\rm Komar}) = (0,0)$ (left part of the curve, i.e. located before the maximum).
The algorithm no longer converges when the maximum value is reached.
From then on,~$\bar{\Omega}_{\mathcal{H}}$ should be lowered to explore more and more rapidly rotating solutions (right part of the curve).
But numerically, using the ``maximum'' solution as initial guess to reach a solution with a smaller value of~$\bar{\Omega}_{\mathcal{H}}$ will actually yield the less rapidly rotating solution (i.e. going backward on the left part of the curve) rather than the more rapidly rotating solution that has the same dimensionless angular velocity~$\bar{\Omega}_{\mathcal{H}}$ but located to the right of the maximum.

Finding a way to ``jump'' over the maximum in order to explore the right part of the curve is a nontrivial issue: one must use another quantity, easily handled numerically, which does parametrize the black hole solutions in a bijective way at least in a neighborhood of the maximum, unlike~$\bar{\Omega}_{\mathcal{H}}$.
Attempts using the dimensionless surface gravity (discussed in the following section) and other parameters fulfilling this condition were unsuccessful so far.
This is why the highest points marked on Fig.~\ref{fig_adOm_vs_adJ} for~$\bar{\gamma} = 0.12$ and~$0.18$ represent the last solutions that could be reached, beyond which the numerical algorithm does not converge anymore, revealing the proximity of a maximum value.

    \subsection{Surface gravity}
    \label{section_surface_gravity}

In a circular spacetime, the zeroth law of black hole mechanics holds \cite{heusler_uniqueness_book,Carter_Houches}, i.e. the surface gravity is homogeneous on the horizons of stationary black holes.
To check this for the solutions presented here, the dimensionless quantity~$\bar{\kappa}$ corresponding to surface gravity~$\kappa$ was extracted according to the following formula:
\begin{eqnarray}
\label{eq_adkappa}
\bar{\kappa} \equiv r_{\mathcal{H}} \kappa = \frac{1}{A} \partial_{\bar{r}}N_{|_{1}}.
\end{eqnarray}

In all solutions, the relative variations of~$\bar{\kappa}$ on the horizon are smaller than~$10^{-6}$, confirming that the surface gravity is numerically homogeneous on the horizon.

\begin{figure}
    \begin{center}
    \includegraphics[width=0.5\textwidth]{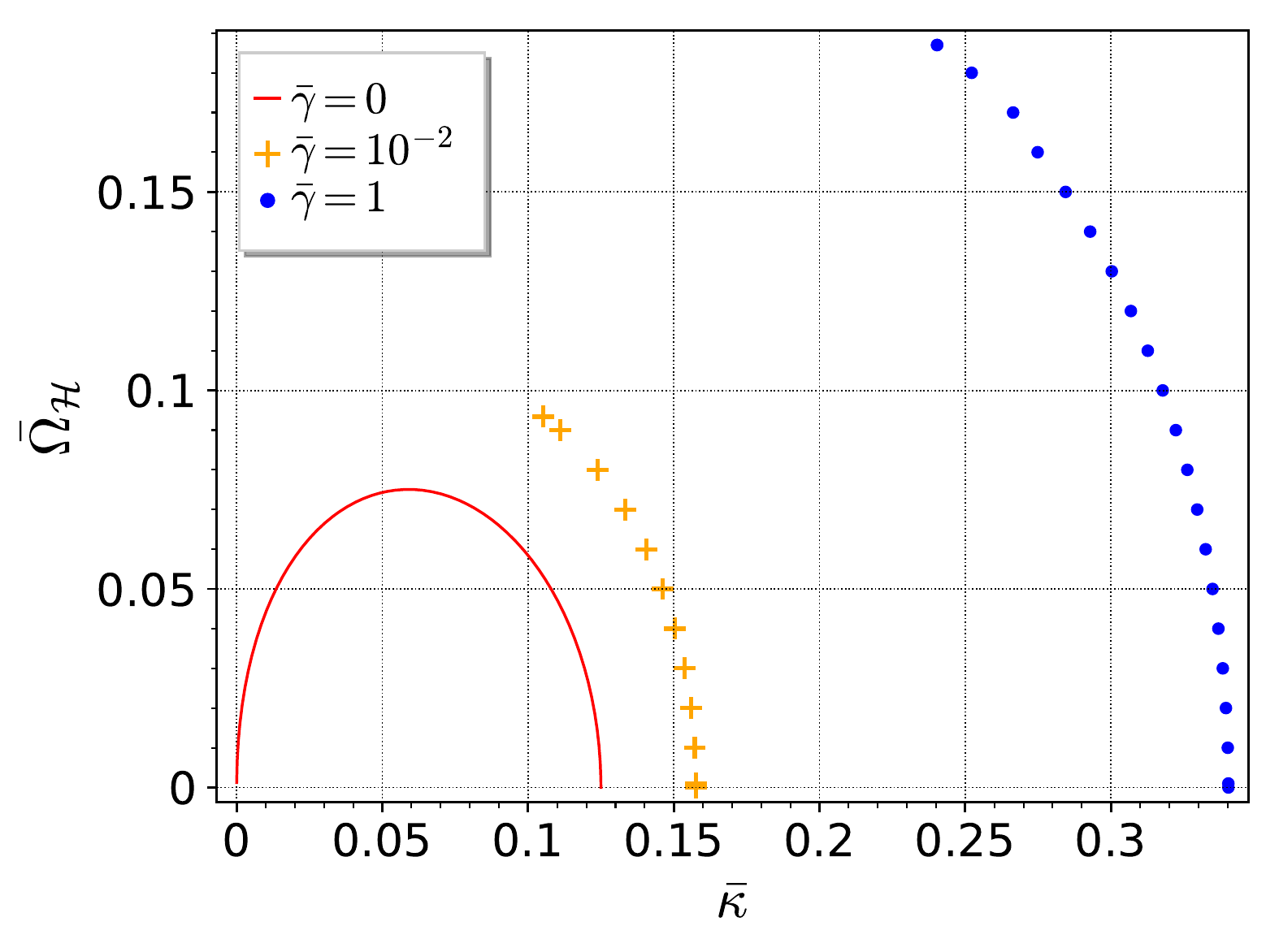}
    \end{center}
\caption{Angular velocity with respect to surface gravity.}
\label{fig_adOm_vs_adKappa}
\end{figure}

The relation between~$\bar{\kappa}$ and~$\bar{\Omega}_{\mathcal{H}}$ is represented on Fig.~\ref{fig_adOm_vs_adKappa}.
For each~$\bar{\gamma}$, the static and spherically symmetric case corresponds to the only point such that~$\bar{\Omega}_{\mathcal{H}} = 0$ but~$\bar{\kappa} \neq 0$, while the origin of the graph, i.e.~$(\bar{\Omega}_{\mathcal{H}},\bar{\kappa}) = (0,0)$ corresponds to extremal cases.
The explicit case of Kerr is again represented by a solid red curve for comparison with GR.

    \subsection{Ergoregion}
    \label{section_ergoregion}

\begin{figure}
    \begin{center}
        \subfloat[Ergoregions of the cubic Galileon solutions.]
        {
            \includegraphics[clip, trim=0cm 0cm 0cm 4.5cm, width=0.45\textwidth]{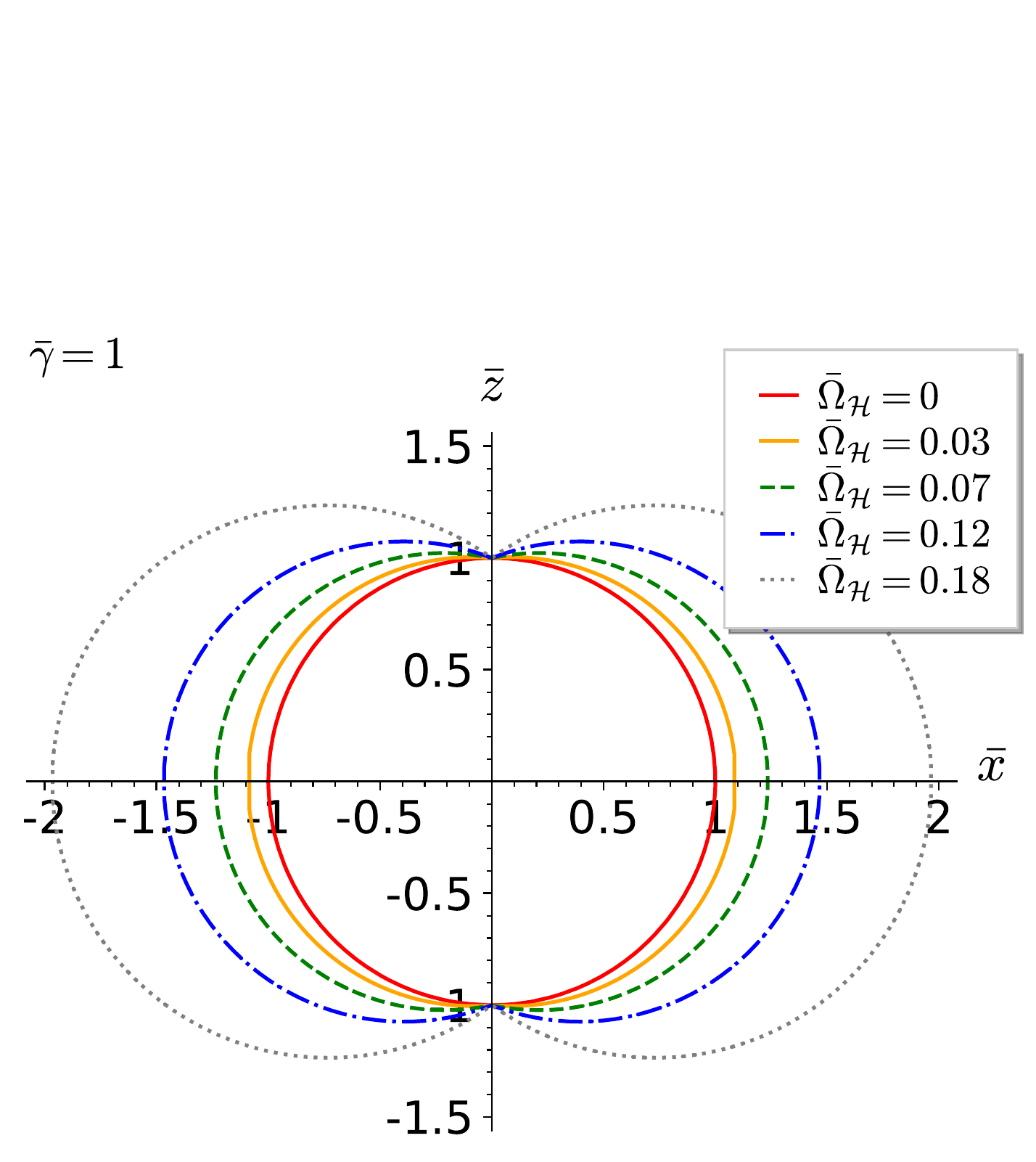}
            \label{fig_rot_ergo_increase_adOm}
        }
        \subfloat[Comparison with Kerr.]
        {
            \includegraphics[clip, trim=0cm 0cm 4.5cm 4.5cm, width=0.55\textwidth]{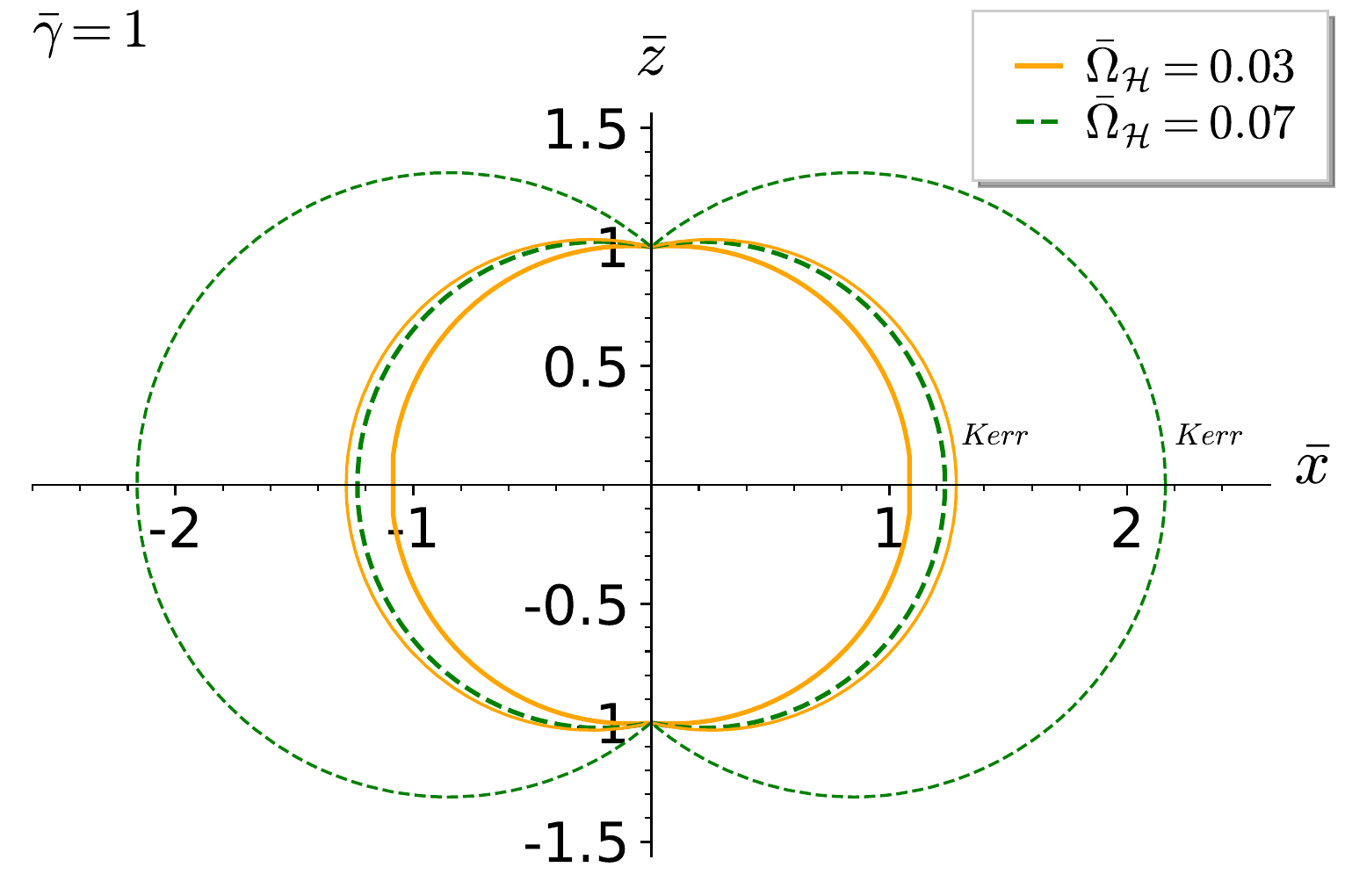}
            \label{fig_rot_ergo_increase_adOm_vs_Kerr}
        }
    \end{center}
\caption{Location of the ergoregion.}
\label{fig_rot_ergo}
\end{figure}

Locating the ergoregions of the rotating solutions provides another evidence of deviations from GR.
Figure~\ref{fig_rot_ergo_increase_adOm} displays the ergoregions corresponding to various angular velocities~$\bar{\Omega}_{\mathcal{H}}$ at fixed coupling~$\bar{\gamma} = 1$ and Fig.~\ref{fig_rot_ergo_increase_adOm_vs_Kerr} compares two of them with Kerr (same color meaning same angular velocity).
On both figures, the ergoregions are plotted in terms of Cartesian-like coordinates yet based on the quasi-isotropic coordinates: $(\bar{x},\bar{z}) = (\bar{r}\sin\theta, \bar{r}\cos\theta)$.
This explains the irregularities observed at the poles even in the case of Kerr, although none is observed in the familiar Boyer-Lindquist coordinates: the change of coordinates from Boyer-Lindquist to quasi-isotropic coordinates is not regular at the poles.

The ergoregions of the cubic Galileon solutions generically have the same shape as Kerr: they coincide with the horizon at the poles and get thicker towards the equator.
They grow as~$\bar{\Omega}_{\mathcal{H}}$ increases, yet they are thinner than Kerr for a given angular velocity.

\section{Conclusions}

Numerical configurations describing asymptotically flat hairy rotating black holes in the cubic Galileon theory have been presented.
They are based on a scalar ansatz involving a linear time-dependence and a circular approximation of the metric.
To realize asymptotic flatness, these Galileon solutions correspond to the special case of vanishing bare cosmological constant and kinetic coupling; they are thus dominated by the DGP term~$(\partial \phi)^{2} \Box\phi$.
The remaining coupling~$\gamma$ induces significant deviations from the Kerr metric on different physical quantities such as surface gravity and angular momentum.
In addition, these asymptotically flat solutions feature convergence towards Minkowski faster than Schwarzschild, which can be understood as a vanishing Komar mass at infinity.

Extreme angular velocities (and possibly extremal cases) were not reached yet but could be handled in future work, along with the search for asymptotically (anti-)de Sitter solutions (meaning nonzero~$\Lambda$ and/or~$\eta$) and the integration of the null and timelike geodesics around such black holes. The key to approach the first problem would be to find an initial guess for rapid rotation. One possible approach would be to take a Kerr background with a scalar stemming from the geodesic analogy similar to \cite{Charmousis:2019vnf}.
Investigation on the latter point would allow to determine whether closed orbits are possible (and up to what distance to the black hole) in spite of the non-Schwarzschild asymptotics.
Integration of the null geodesics would simulate the astrophysical imaging of an emitting accretion torus surrounding the Galileon black holes, to be compared with results obtained for other types of compact objects \cite{image_Kerr_scalar_hair,image_boson,image_regular_BH}.
Such investigations now have a clear astrophysical relevance in regards of the observations from GRAVITY \cite{GRAVITY_redshift_S2, GRAVITY_motion_ISCO} and the Event Horizon Telescope \cite{EHT_Shape_SgrA, EHT_Shadow_M87} and we hope to be reporting on these issues in the near future.

\section*{Acknowledgements}

The authors thank Gilles Esposito-Far\`ese and Eugeny Babichev for several helpful discussions.
The authors acknowledge valuable support from the CNRS project 80PRIME-TNENGRAV.

\appendix

\section{No-scalar-hair theorem for the cubic Galileon}
\label{appdx_no_hair}

A static and spherically symmetric spacetime admits coordinates~$(t,R,\theta,\varphi)$ with respect to which the metric can be written as~(\ref{eq_metric_Schw_like}).

If the Galileon field features the same symmetries, it only depends on the radial coordinate~$R$, and the~$(tR)$ metric equation (i.e. equation~(\ref{eq_4_1}) in which~$q$ is set to~$0$) reads
\begin{eqnarray}
\label{eq_tR_no_q}
\phi' \left[ f \phi' \left( \frac{h'}{h} + \frac{4}{R} \right) - \frac{2\eta}{\gamma} \right] = 0.
\end{eqnarray}

The general no-hair theorem \cite{no_hair_Galileon,slowly_rotating_no_hair} assumes the Galileon Lagrangian to contain a standard kinetic term, i.e.~$\eta \neq 0$.
Yet, for the cubic Galileon, the case~$\eta = 0$ can be included in the theorem, or yields a nontrivial hairy solution if asymptotic flatness is abandoned (see below).

\paragraph{Case~$\eta \neq 0$}
The metric equation~(\ref{eq_4_2}) in which~$q$ is set to~$0$ gives
\begin{eqnarray}
\phi'^{2} = -\frac{2\zeta}{\eta} \left[ \frac{h'}{Rh} + \frac{1}{f} \left( \frac{f-1}{R^{2}} + \Lambda \right) \right].
\end{eqnarray}

Then, the asymptotic flatness requirements~(\ref{eq_asympt_flat_h})-(\ref{eq_asympt_flat_f}) imply that~$\phi'^{2} \longrightarrow - 2 \zeta \Lambda/\eta$.
In particular,~$\phi'$ is bounded at infinity, so that
\begin{eqnarray}
f \phi' \left( \frac{h'}{h} + \frac{4}{R} \right) \longrightarrow 0.
\end{eqnarray}

If the latter term was nonzero at some point,
its absolute value would get smaller than e.g.~$\eta/\gamma$ at some other point while remaining strictly positive, which would require~$\phi' \neq 0$.
This would contradict~(\ref{eq_tR_no_q}), in which one could simplify the overall factor~$\phi'$ while having no chance for~$f \phi' (h'/h + 4/R) = 2\eta/\gamma$ to hold.

Therefore,~$f \phi' (h'/h + 4/R)$ must vanish everywhere and equation~(\ref{eq_tR_no_q}) finally implies that~$\phi$ is trivial (up to a meaningless constant shift).

\paragraph{Case~$\eta = 0$}
A hairy solution would feature nonzero~$\phi'$ on some interval~$I$, which can be assumed to either extend to infinity, or to be such that~$\phi'$ is zero beyond some upper bound.
According to~(\ref{eq_tR_no_q}) with~$\eta = 0$, one would have
\begin{eqnarray}
\label{eq_h_q0_eta0}
h = \frac{h_{1}}{R^{4}} \text{ over } I,
\end{eqnarray}
where~$h_{1}$ is an integration constant (whose sign must be the same as~$f$ on~$I$ for the metric to be Lorentzian).
Yet the expression~(\ref{eq_h_q0_eta0}) does not meet with the asymptotic behaviour~(\ref{eq_asympt_flat_h}) so that~$I$ cannot extend to infinity.
This can also be seen from the metric equation~(\ref{eq_4_2}) in which~$\eta$ is set to~$0$:
\begin{eqnarray}
\label{eq_f_q0_eta0_Lambda0}
f = \left( \frac{1}{R^{2}} - \Lambda \right) \left( \frac{h'}{Rh} + \frac{1}{R^{2}} \right)^{-1} = \frac{\Lambda R^{2} - 1}{3} \text{ over } I,
\end{eqnarray}
which does not meet with the asymptotic behaviour~(\ref{eq_asympt_flat_f}) either.

Therefore~$\phi'$ should vanish at some point~$R_{0}$ and remain zero up to infinity;
whether this is possible to realize in a smooth way or not relies on the equation~(\ref{eq_scalar_eta0_nonflat}) which provides the expression of~$\phi'$ on~$I$.
But anyway, beyond~$R_{0}$, $h$ and~$f$ would become Schwarzschild, with no chance to match~(\ref{eq_h_q0_eta0}) and~(\ref{eq_f_q0_eta0_Lambda0}) at~$R_{0}$ in a smooth way:
\begin{eqnarray}
h = f = 1 - \frac{4R_{0}}{3R},\ R \geq R_{0},
\end{eqnarray}
so that only the Schwarzschild behaviour outside the event horizon located at~$R = 4R_{0}/3$ and a trivial Galileon remain meaningful.

\paragraph{Solutions with~$\eta = 0$}
Abandoning asymptotic flatness allows us to use the expressions~(\ref{eq_h_q0_eta0}) and~(\ref{eq_f_q0_eta0_Lambda0}) everywhere up to infinity, and thus inject them into equation~(\ref{eq_4_3}) in which~$q$ and~$\eta$ are set to~$0$.
The resulting equation takes the form
\begin{eqnarray}
\left[ \left( R^{2} \sqrt{fh} \phi' \right)^{3} \right]' =
\frac{2\zeta}{\gamma} G_{\Lambda}',
\end{eqnarray}
where
\begin{eqnarray}
G_{\Lambda}' = \left( \frac{2}{R^{2}} - 3 \Lambda \right) \sqrt{ \frac{3 h_{1}^{3}}{\Lambda R^{2} - 1} },
\end{eqnarray}
which integrates into
\vspace{-1cm}
\begin{spacing}{1.8}
\begin{eqnarray}
\hspace{-2.5cm}
G_{\Lambda} \hspace{-1mm} = \hspace{-1mm}
\left\{ \hspace{-2mm}
\begin{array}{l}
\sqrt{3 h_{1}^{3}} \left[ \frac{2\sqrt{\Lambda R^{2} - 1}}{R} - 3 \sqrt{\Lambda} \text{arcosh}\left( \sqrt{\Lambda} R \right) \right] \text{ if~$\Lambda > 0$, $R > \frac{1}{\sqrt{\Lambda}}$ and hence~$h_{1} > 0$}, \\
- \sqrt{3 \vert h_{1} \vert^{3}} \left[ \frac{2\sqrt{ 1-\Lambda R^{2} }}{R} + 3 \sqrt{\Lambda} \text{arcsin}\left( \sqrt{\Lambda} R \right) \right] \text{ if~$\Lambda > 0$, $R < \frac{1}{\sqrt{\Lambda}}$ and hence~$h_{1} < 0$}, \\
\sqrt{3 \vert h_{1} \vert^{3}} \left[ - \frac{2\sqrt{ 1-\Lambda R^{2} }}{R} + 3 \sqrt{\vert\Lambda\vert} \text{arsinh}\left( \sqrt{\vert\Lambda\vert} R \right) \right] \text{ if~$\Lambda \leq 0$ and hence~$h_{1} < 0$}.
\end{array}
\right.
\end{eqnarray}
\end{spacing}

In any case, one finally has
\begin{eqnarray}
\label{eq_scalar_eta0_nonflat}
\phi' = \sqrt{ \frac{3}{h_{1} \left( \Lambda R^{2} - 1 \right) } } \left( \frac{2\gamma}{\zeta} G_{\Lambda} + \alpha \right)^{1/3},
\end{eqnarray}
where~$\alpha$ is an integration constant.

If~$\Lambda \leq 0$, then~$t$ is a spacelike coordinate and~$R$ is timelike, so that the expressions~(\ref{eq_h_q0_eta0}),~(\ref{eq_f_q0_eta0_Lambda0}) and~(\ref{eq_scalar_eta0_nonflat}) describe a time-dependent metric and a homogeneous, time-dependent scalar field.
It is also the case if~$\Lambda > 0$ and~$R < 1/\sqrt{\Lambda}$, so that the time coordinate~$R$ is bounded, like the interior Schwarzschild solution.
Finally, if~$\Lambda > 0$, the expressions~(\ref{eq_h_q0_eta0}),~(\ref{eq_f_q0_eta0_Lambda0}) and (\ref{eq_scalar_eta0_nonflat}) describe the exterior domain of a hairy black hole spacetime with an event horizon located at~$R = 1/\sqrt{\Lambda}$.
Asymptotically,~$\phi'$ converges to zero as~$\ln(R)/R$.

\section{Source terms and scalar equation}
\label{appdx_RHS}

The explicit expressions of the source terms and the scalar equation exposed below are justified in a Jupyter notebook based on the free software SageMath\footnote{\url{https://www.sagemath.org}, \url{https://sagemanifolds.obspm.fr}}.
The notebook is available at the following url: \url{https://share.cocalc.com/share/6cfa5f27-1564-4bd8-9b0c-fcb3c7d0f325/2019-09-29-155358/metric_and_scalar_equations_cubic_Galileon.ipynb?viewer=share}.

In the explicit expressions, the following notations are used for any functions~$f$, $g$ and~$h$ of~$\bar{r}$ and~$\theta$:
\begin{eqnarray}
\partial f \partial g = \partial_{\bar{r}}f \, \partial_{\bar{r}}g + \frac{1}{\bar{r}^{2}} \partial_{\theta}f \, \partial_{\theta}g,
\\
\mathcal{H}^{(0)}_{f}[g,h] =
    \left(
    \begin{array}{c}
        \partial_{\bar{r}}g \\
        \frac{1}{\bar{r}} \partial_{\theta}g
    \end{array}
    \right)^{T}
    \left(
    \begin{array}{l r}
        \partial^{2}_{\bar{r}\bar{r}}f                       & \frac{1}{\bar{r}} \partial^{2}_{\bar{r}\theta}f    \\
        \frac{1}{\bar{r}} \partial^{2}_{\bar{r}\theta}f      & \frac{1}{\bar{r}^{2}} \partial^{2}_{\theta\theta}f
    \end{array}
    \right)
    \left(
    \begin{array}{c}
        \partial_{\bar{r}}h \\
        \frac{1}{\bar{r}} \partial_{\theta}h
    \end{array}
    \right),
\\
\mathcal{H}^{(1)}_{f}[g,h] =
    \left(
    \begin{array}{c}
        \frac{1}{\bar{r}} \partial_{\theta}g \\
        - \partial_{\bar{r}}g
    \end{array}
    \right)^{T}
    \left(
    \begin{array}{l r}
        \partial^{2}_{\bar{r}\bar{r}}f                       & \frac{1}{\bar{r}} \partial^{2}_{\bar{r}\theta}f    \\
        \frac{1}{\bar{r}} \partial^{2}_{\bar{r}\theta}f      & \frac{1}{\bar{r}^{2}} \partial^{2}_{\theta\theta}f
    \end{array}
    \right)
    \left(
    \begin{array}{c}
        \frac{1}{\bar{r}} \partial_{\theta}h \\
        - \partial_{\bar{r}}h
    \end{array}
    \right),
\\
\mathcal{H}^{(2)}_{f}[g,h] =
    \left(
    \begin{array}{c}
        \partial_{\bar{r}}g \\
        \frac{1}{\bar{r}} \partial_{\theta}g
    \end{array}
    \right)^{T}
    \left(
    \begin{array}{l r}
        \partial^{2}_{\bar{r}\bar{r}}f                       & \frac{2}{\bar{r}} \partial^{2}_{\bar{r}\theta}f    \\
        \frac{2}{\bar{r}} \partial^{2}_{\bar{r}\theta}f      & \frac{1}{\bar{r}^{2}} \partial^{2}_{\theta\theta}f
    \end{array}
    \right)
    \left(
    \begin{array}{c}
        \partial_{\bar{r}}h \\
        \frac{1}{\bar{r}} \partial_{\theta}h
    \end{array}
    \right).
\end{eqnarray}

Then, the right-hand side terms of equations~(\ref{eq_metric_QI_N_bis})-(\ref{eq_metric_QI_adom_bis}) read

\begin{eqnarray}
\label{eq_RHS_N}
\mathcal{S}_{N} =
&\frac{N \left( B\bar{r}\sin\theta \right)^{2}}{2} \partial\bar{\omega} \partial\bar{\omega}
    - \frac{N^{2}}{B} \partial N \partial B
    - N A^{2} \left( \bar{\eta} + \bar{\Lambda} N^{2} \right) \nonumber\\
&- \frac{\bar{\gamma}}{2} \left(1 + \frac{N^{2} \partial\bar{\Psi}\partial\bar{\Psi}}{A^{2}} \right)
    \left( N \Delta_{3}\bar{\Psi} + \partial\bar{\Psi}\partial N + \frac{N \partial\bar{\Psi}\partial B}{B} \right),
\end{eqnarray}

\begin{eqnarray}
\label{eq_RHS_NA}
\mathcal{S}_{A} =
&\frac{N^4}{A}\partial A \partial A
    + 2 N^3 \partial A \partial N
    + \frac{3 A (N B\bar{r}\sin\theta)^{2}}{4} \partial\bar{\omega} \partial\bar{\omega} \nonumber \\
& + \frac{\bar{\eta} N^{2} A}{2} \left( N^{2} \partial\bar{\Psi}\partial\bar{\Psi} - A^{2} \right)
    - \bar{\Lambda} A^{3} N^{4} \nonumber \\
&- \bar{\gamma} \left(
    N A \partial\bar{\Psi}\partial N
    - \frac{N^{4} \partial\bar{\Psi}\partial\bar{\Psi} \ \partial\bar{\Psi}\partial A}{A^{2}} \right. \nonumber \\
& \hspace{1.2cm} \left. + \frac{1}{A}   \left[
    N^{4}\mathcal{H}^{(0)}_{\bar{\Psi}}[\bar{\Psi},\bar{\Psi}]
    - \frac{N^{4} \partial_{r}\bar{\Psi}}{\bar{r}^{3}} \left(\partial_{\theta}\bar{\Psi}\right)^{2}
    \right]
    \right),
\end{eqnarray}

\begin{eqnarray}
\label{eq_RHS_NB}
\mathcal{S}_{B} =
&- B \bar{r} \sin\theta \left[ N A^{2} \left( \bar{\eta} + 2 \bar{\Lambda} N^{2} \right)  \right. \nonumber \\ %] to close \left opening square bracket not to spoil coloration
&  \hspace{2.2cm} + \left. \frac{\bar{\gamma} N^2 \partial\bar{\Psi}\partial\bar{\Psi}}{A^{2}}
    \left( N \Delta_{3}\bar{\Psi} + \partial\bar{\Psi}\partial N  + \frac{N \partial\bar{\Psi}\partial B}{B} \right) \right],
\end{eqnarray}

\begin{eqnarray}
\label{eq_RHS_adom}
\mathcal{S}_{\bar{\omega}} =
\frac{N \bar{\omega}}{\bar{r} \sin\theta}
+ \bar{r} \sin\theta \left( \partial\bar{\omega} \partial N
- \frac{3 N}{B} \partial \bar{\omega} \partial B \right)
\end{eqnarray}
and the scalar equation takes the form
\begin{eqnarray}
\hspace{-2.5cm} 0 =
& \hspace{-1.8cm} - \bar{\eta} N^{3} A^{2} \left( N\Delta_{3}\bar{\Psi} + \partial\bar{\Psi} \partial N + \frac{N \partial\bar{\Psi} \partial B}{B} \right)    \nonumber\\
& \hspace{-1.8cm} + \bar{\gamma} \left\{ A^{2} \left( N\Delta_{3}N + \frac{N}{B} \partial N \partial B - 2 \partial N \partial N \right)             \right.  \nonumber\\
& \hspace{-0.7cm} + 2 N \left( N\Delta_{3}\bar{\Psi} + \partial\bar{\Psi} \partial N + \frac{N \partial\bar{\Psi} \partial B}{B} \right)
    \left( \frac{N^{2} \partial\bar{\Psi}\partial A}{A} - N \partial\bar{\Psi}\partial N \right)                                \nonumber\\
& \hspace{-0.7cm} - 2 \left( N^{2} \Delta_{2}\bar{\Psi} + \frac{N^{2} \partial\bar{\Psi} \partial B}{B} \right)
     \left( N^{2} \Delta_{3}\bar{\Psi} - \frac{N^{2} \partial_{\bar{r}}\bar{\Psi}}{\bar{r}} \right)                                                \nonumber\\
& \hspace{-0.7cm} + \frac{2 N^{2}}{\bar{r}^{2}} \partial^{2}_{\theta\theta}\bar{\Psi} \left( N^{2} \Delta_{2}\bar{\Psi} - \frac{N^{2} \partial\bar{\Psi} \partial A}{A} \right)  \nonumber\\
& \hspace{-0.7cm} - N^{3} \partial\bar{\Psi} \partial\bar{\Psi}
    \left(\frac{N}{A} \left[ \Delta_{3}A - \frac{4}{\bar{r}}\partial_{\bar{r}}A \right]
        + \frac{N}{B} \Delta_{2}B
        + \frac{\partial N \partial A}{A}
        - \frac{3N}{A^{2}} \partial A \partial A
        + \frac{N}{AB} \partial A \partial B
    \right)              \nonumber\\
& \hspace{-0.7cm} + 2 \left( N \partial\bar{\Psi} \partial N \right)^{2}                                                         \nonumber\\
& \hspace{-0.7cm} - N^{3} \, \mathcal{H}^{(0)}_{N}[\bar{\Psi},\bar{\Psi}]                              \nonumber\\
& \hspace{-0.7cm} + \frac{N^{4} \mathcal{H}^{(1)}_{B}[\bar{\Psi},\bar{\Psi}]}{B}           \nonumber\\
& \hspace{-0.7cm} - \frac{2N^{4} \mathcal{H}^{(2)}_{\bar{\Psi}}[\bar{\Psi},A]}{A}                               \nonumber\\
& \hspace{-0.7cm} + 2 \left( \left[ N^{2} \partial^{2}_{\bar{r}\bar{r}}\bar{\Psi} \right]^{2} +
            \left[ \frac{N^{2}}{\bar{r}^{2}}\partial_{\theta}\bar{\Psi} - \frac{N^{2} \partial_{\bar{r}\theta}\bar{\Psi}}{\bar{r}} \right]^{2}       \right)               \nonumber\\
& \hspace{-0.7cm} - \frac{2 N^{2} \partial_{\bar{r}}\bar{\Psi} \partial_{\bar{r}} A}{A}
    \left( N^{2} \partial^{2}_{\bar{r}\bar{r}}\bar{\Psi} + \frac{2 N^{2} \partial_{\bar{r}}\bar{\Psi}}{\bar{r}} - \frac{N^{2}}{\bar{r}^{2}} \partial^{2}_{\theta\theta}\bar{\Psi} \right)                 \nonumber\\
& \hspace{-0.7cm} - \frac{N^{4} (\partial_{\bar{r}}\bar{\Psi})^{2}}{\bar{r}} \frac{ \partial_{\bar{r}} B}{B}                                                          \nonumber\\
& \hspace{-0.7cm} + \left. \frac{N^{2}}{\bar{r}^{3}} \partial_{\theta}\bar{\Psi}
    \left( 2 N \partial_{\bar{r}}\bar{\Psi} \partial_{\theta}N - N \partial_{\theta}\bar{\Psi} \partial_{\bar{r}}N \right)
\right\}.
\end{eqnarray}

The function~$N$ could be factored out in many places but instead it is explicitly left everywhere it is needed to counterbalance divergences on the horizon.
More precisely, it appears as a factor in front of all the quantities that involve the radial derivative of~$\bar{\Psi}$, in order to form terms that remain finite on the horizon.

\section*{References}
\newcommand{\eprint}[2][]{\href{https://arxiv.org/abs/#2}{arXiv:#2}}
\bibliographystyle{iopart-num}
\bibliography{v2}

\end{document}